\documentclass[onecolumn]{emulateapj}
\shorttitle{Estimating QSO Mass Functions}
\shortauthors{Kelly et al.}

\begin{document}
  
  \title{Determining Quasar Black Hole Mass Functions from their Broad
  Emission Lines: Application to the Bright Quasar Survey}

  \author{Brandon C. Kelly\altaffilmark{1,2,3,4},
    Marianne Vestergaard\altaffilmark{5}, Xiaohui Fan\altaffilmark{3}}

  \altaffiltext{1}{bckelly@cfa.harvard.edu}
  \altaffiltext{2}{Hubble Fellow}
  \altaffiltext{3}{Department of Astronomy, University of Arizona, Tucson, AZ 85721}
  \altaffiltext{4}{Harvard-Smithsonian Center for Astrophysics, 
      60 Garden St, Cambridge, MA 02138}
  \altaffiltext{5}{Department of Physics and Astronomy, Tufts University, 
    Medford, MA 02155}

  \begin{abstract}
    We describe a Bayesian approach to estimating quasar black hole
    mass functions (BHMF) when using the broad emission lines to
    estimate black hole mass. We show how using the broad line mass
    estimates in combination with statistical techniques developed for
    luminosity function estimation (e.g., the $1 / V_a$ correction)
    leads to statistically biased results. We derive the likelihood
    function for the BHMF based on the broad line mass estimates, and
    derive the posterior distribution for the BHMF, given the observed
    data. We develop our statistical approach for a flexible model
    where the BHMF is modelled as a mixture of Gaussian
    functions. Statistical inference is performed using markov chain
    monte carlo (MCMC) methods, and we describe a Metropolis-Hasting
    algorithm to perform the MCMC. The MCMC simulates random draws
    from the probability distribution of the BHMF parameters, given
    the data, and we use a simulated data set to show how these random
    draws may be used to estimate the probability distribution for the
    BHMF. In addition, we show how the MCMC output may be used to
    estimate the probability distribution of any quantities derived
    from the BHMF, such as the peak in the space density of
    quasars. Our method has the advantage that it is able to
    constrain the BHMF even beyond the survey detection limits at the
    adopted confidence level, accounts for measurement errors and the
    intrinsic uncertainty in broad line mass estimates, and provides a
    natural way of estimating the probability distribution of any
    quantities derived from the BHMF. We conclude by using our method
    to estimate the local active BHMF using the $z < 0.5$ Bright
    Quasar Survey sources. At $z \sim 0.2$, the quasar BHMF falls off
    approximately as a power law with slope $\sim 2$ for $M_{BH}
    \gtrsim 10^8 M_{\odot}$. Our analysis implies that at a given
    $M_{BH}$, $z < 0.5$ broad line quasars have a typical Eddington
    ratio of $\sim 0.4$ and a dispersion in Eddington ratio of
    $\lesssim 0.5$ dex.
  \end{abstract}
  
  
  \section{INTRODUCTION}

  It is widely accepted that the extraordinary activity associated
  with quasars\footnote[1]{Throughout this work we will use the terms
  quasar and AGN to refer generically to broad line AGNs. No
  luminosity difference between the two is assumed.} involves
  accretion onto a supermassive black hole (SMBH). The correlation
  between SMBH mass and both host galaxy luminosity
  \citep[e.g.,][]{korm95,mag98,mclure01,marc03} and stellar velocity
  dispersion \citep[$M_{BH}$--$\sigma$ relationship,
  e.g.,][]{gebhardt00, merritt01, trem02}, together with the fact that
  quasars have been observed to reside in early-type galaxies
  \citep{mclure99,kuk01,mcleod01,nolan01,perc01,dunlop03}, implies
  that the evolution of spheroidal galaxies and quasars is intricately
  tied together
  \citep[e.g.,][]{silk98,haehn00,merr04,dimatt05,hopkins06}.
  Therefore, investigating the evolution of active super-massive black
  holes (SMBHs) is an important task of modern astronomy, giving
  insight into the importance of AGN activity on the formation of
  structure in the universe. Determination of the comoving number
  density, energy density, and mass density of active black holes is a
  powerful probe of the quasar-galaxy connection and the evolution of
  active black holes.

  Recently, advances in reverberation mapping \citep[e.g.,][]{peter04}
  have made it possible to estimate the masses of black holes for
  broad line AGN. A correlation has been found between the size of the
  region emitting the broad lines and the luminosity of the AGN
  \citep{kaspi05,bentz06}, allowing one to use the source luminosity
  to estimate the distance between the broad line region (BLR) and the
  central black hole. In addition, one can estimate the velocity
  dispersion of the BLR gas from the broad emission line width. One
  then combines the BLR size estimate with the velocity estimate to
  obtain a virial black hole mass as $M_{BH} \propto L^{b} V^2$, where
  $b \approx 1/2$
  \citep[e.g.,][]{wandel99,mclure02,vest02,vest06}. Estimates of
  $M_{BH}$ obtained from the broad emission lines have been used to
  estimate the distribution of quasar black hole masses at a variety
  of redshifts
  \citep[e.g.,][]{mclure04,vest04,koll06,wang06,greene07,vest08,fine08}.

  Given the importance of the BHMF as an observational constraint on
  models of quasar evolution, it is essential that a statistically
  accurate approach be employed when estimating the BHMF. However, the
  existence of complicated selection functions hinders this. A variety
  of methods have been used to accurately account for the selection
  function when estimating the quasar luminosity function. These
  include various binning methods
  \citep[e.g.,][]{schmidt68,avni80,page00}, maximum-likelihood fitting
  \citep[e.g.,][]{marshall83,fan01}, a semi-parameteric approach
  \citep{schafer07}, and Bayesian approaches \citep[e.g.,][hereafter
    KFV08]{andreon05,kfv08}. In addition, there have been a variety of methods
  proposed for estimating the cumulative distribution function of the
  luminosity function \citep[e.g.,][]{lynden71,efron92,mal99}. While
  these techniques have been effective for estimating luminosity
  functions, estimating the BHMF from the broad line mass estimates is
  a more difficult problem, and currently there does not exist a
  statistically correct method of estimating the BHMF.

  If we could directly measure black hole mass for quasars, and if the
  selection function only depended on $M_{BH}$ and $z$, then we could
  simply employ the formalism developed for luminosity function
  estimation, after replacing $L$ with $M_{BH}$. However, surveys are
  selected based on luminosity and redshift, not on $M_{BH}$. At any
  given luminosity there exists a range in black hole mass, and thus
  one cannot simply employ the luminosity selection function `as-is'
  to correct for the flux limit. In other words, completeness in flux
  is not the same thing as completeness in $M_{BH}$, and the use of a
  flux selection results in a softer selection function for
  $M_{BH}$. Moreover, we cannot directly observe $M_{BH}$ for large
  samples of quasars, but rather derive an estimate of $M_{BH}$ from
  their broad emission lines. The intrinsic uncertainty on $M_{BH}$
  derived from the broad emission lines is $\sim 0.4$ dex
  \citep{vest06}, and the uncertainty on $M_{BH}$ broadens the
  inferred distribution of $M_{BH}$
  \citep[e.g.,][]{kelly07b,shen07,fine08}. As a result, even if there
  is no flux limit, the BHMF inferred directly from the broad line
  mass estimates will be systematically underestimated near the peak
  and overestimated in the tails. In order to ensure an accurate
  estimate of the BHMF it is important to correct for the uncertainty
  in the estimates of $M_{BH}$.

  Motivated by these issues, we have developed a Bayesian method for
  estimating the BHMF. In KFV08 we derived the likelihood function and
  posterior probability distribution for luminosity function
  estimation, and we described a mixture of Gaussian functions model
  for the luminosity function. In this work, we extend our statistical
  method and derive the likelihood function of the BHMF by relating
  the observed data to the true BHMF, and derive the posterior
  probability distribution of the BHMF parameters, given the observed
  data. While the likelihood function and posterior are valid for any
  parameteric form, we focus on a flexible parameteric model where the
  BHMF is modeled as a sum of Gaussian functions. This is a type of
  `non-parameteric' approach, where the basic idea is that the
  individual Gaussian functions do not have any physical meaning, but
  that given enough Gaussian functions one can obtain a suitably
  accurate approximation to the true BHMF. Modeling the BHMF as a
  mixture of normals avoids the problem of choosing a particular
  parameteric form, especially in the absence of any guidance from
  astrophysical theory. In addition, we describe a markov chain monte
  carlo (MCMC) algorithm for obtaining random draws from the posterior
  distribution. These random draws allow one to estimate the posterior
  distribution for the BHMF, as well as any quantities derived from
  it. The MCMC method therefore allows a straight-forward method of
  calculating errors on any quantity derived from the BHMF. Because
  the Bayesian approach is valid for any sample size, one is able to
  place reliable constraints on the BHMF and related quantities, even
  where the survey becomes incomplete.

  Because of the diversity and mathematical complexity of some parts
  of this paper, we summarize the main results here. We do this so
  that the reader who is only interested in specific aspects of this
  paper can conveniently consult the sections of interest.
  \begin{itemize}
  \item
    In \S~\ref{s-mbllik} we derive the general form of the likelihood
    function for black hole mass function estimation based on quasar
    broad emission lines. Because we can not directly observe $M_{BH}$
    for a large sample of quasars, the likelihood function gives the
    probability of observing a set of redshifts, luminosities, and
    line widths, given an assumed BHMF. In \S~\ref{s-selfunc} we
    derive the black hole mass selection function, and discuss how the
    differences between the $M_{BH}$ selection function and the
    luminosity selection function affect estimating the BHMF. The
    reader who is interested in the likelihood function of the broad
    line quasar BHMF, or issues regarding correcting for incompleteness
    in $M_{BH}$, should consult this section.
  \item
    In \S~\ref{s-posterior} we describe a Bayesian approach to black
    hole mass function estimation. We build on the likelihood function
    derived in \S~\ref{s-mbllik} to derive the probability
    distribution of the BHMF, given the observed data (i.e., the
    posterior distribution). The reader who is interested in a
    Bayesian approach to BHMF estimation should consult this section.
  \item
    In \S~\ref{s-smodel} we develop a mixture of Gaussian functions
    model for the black hole mass function, deriving the likelihood
    function and posterior distribution for this model. Under this
    model, the BHMF is modelled as a weighted sum of Gaussian
    functions. This model has the advantage that, given a suitably
    large enough number of Gaussian functions, it is flexible enough
    to give an accurate estimate of any smooth and continuous
    BHMF. This allows the model to adapt to the true BHMF, thus
    minimizing the bias that can result when assuming a parameteric
    form for the BHMF. In addition, we also describe our statistical
    model for the distribution of luminosities at a given $M_{BH}$,
    and the distribution of line widths at a given $L$ and
    $M_{BH}$. These two distribution are necessary in order to link
    the BHMF to the observed set of luminosities and line widths. The
    reader who are interested in employing our mixture of Gaussian
    functions model should consult this section.
  \item
    Because of the large number of parameters associated with black
    hole mass function estimation, Bayesian inference is most easily
    performed by obtaining random draws of the BHMF from the posterior
    distribution.  In \S~\ref{s-mha} we describe a Metropolis-Hastings
    algorithm (MHA) for obtaining random draws of the BHMF from the
    posterior distribution, assuming our mixture of Gaussian functions
    model. The reader who is interested in the computational aspects
    of `fitting' the mixture of Gaussian functions model, or who is
    interested in the computational aspects of Bayesian inference for
    the BHMF, should consult this section.
  \item
    In \S~\ref{s-sim} we use simulation to illustrate the
    effectiveness of our Bayesian Gaussian mixture model for black
    hole mass function estimation. We construct a simulated data set
    similar to the Sloan Digital Sky Survey DR3 Quasar Cataloge
    \citep{dr3qsos}. We then use our mixture of Gaussian functions
    model to recover the true BHMF and show that our mixture model is
    able to place reliable constraints on the BHMF over all values of
    $M_{BH}$. In constrast, we show that estimating the BHMF by
    binning up the broad line mass estimates, and applying a simple $1
    / V_a$ correction, systematically biases the inferred BHMF toward
    larger $M_{BH}$. We also illustrate how to use the MHA output to
    constrain any quantity derived from the BHMF, and how to use the
    MHA output to assess the quality of the fit. Finally, we discuss
    difficulties associated with inferring the distribution of
    Eddington ratios. The reader who is interested in assessing the
    effectiveness of our statistical approach, or who is interested in
    using the MHA output for statistical inference on the BHMF, should
    consult this section.
  \item
    In \S~\ref{s-bqs} we use our statistical method to estimate the $z
    < 0.5$ BHMF from the Bright Quasar Survey sources. We also attempt
    to infer the mean and dispersion in the $z < 0.5$ distribution of
    Eddington ratios. The reader who is interested in the scientific
    results regarding our estimated $z < 0.5$ BHMF should consult this
    section.
  \end{itemize}

  We adopt a cosmology based on the the WMAP best-fit parameters
  \citep[$h=0.71, \Omega_m=0.27, \Omega_{\Lambda}=0.73$,][]{wmap}

  \section{THE LIKELIHOOD FUNCTION}

  \label{s-lik}

  \subsection{NOTATION}

  \label{s-notation}

  We use the common statistical notation that an estimate of a
  quantity is denoted by placing a `hat' above it; e.g.,
  $\hat{\theta}$ is an estimate of the true value of the parameter
  $\theta$. We denote a normal distribution with mean $\mu$ and
  variance $\sigma^2$ as $N(\mu, \sigma^2)$, and we denote as $N_p
  (\mu, \Sigma)$ a multivariate normal distribution with $p$-element
  mean vector $\mu$ and $p \times p$ covariance matrix $\Sigma$. If we
  want to explicitly identify the argument of the Gaussian function,
  we use the notation $N(x|\mu,\sigma^2)$, which should be understood
  to be a Gaussian function with mean $\mu$ and variance $\sigma^2$ as
  a function of $x$. We will often use the common statistical notation
  where ``$\sim$'' means ``is drawn from'' or ``is distributed
  as''. This should not be confused with the common usage of
  ``$\sim$'' implying ``similar to''. For example, $x \sim N(\mu,
  \sigma^2)$ states that $x$ is drawn from a normal distribution with
  mean $\mu$ and variance $\sigma^2$, whereas $x \sim 1$ states that
  the value of $x$ is similar to one.

  \subsection{Likelihood Function for the BHMF Estimated from AGN Broad Emission Lines}

  \label{s-mbllik}

  The black hole mass function, denoted as $\phi(M_{BH}, z) dM_{BH}$,
  is the number of sources per comoving volume $V(z)$ with black hole
  masses in the range $M_{BH}, M_{BH} + dM_{BH}$. The black hole mass
  function is related to the probability distribution of $(M_{BH},z)$ by
  \begin{equation}
    p(M_{BH},z) = \frac{1}{N} \phi(M_{BH},z) \frac{dV}{dz},
    \label{eq-phiconvert}
  \end{equation}
  where $N$ is the total number of sources in the universe, and is
  given by the integral of $\phi$ over $M_{BH}$ and $V(z)$.  If we
  assume a parameteric form for $\phi(M_{BH},z)$, with parameters
  $\theta$, we can derive the likelihood function for the observed
  data. The likelihood function is the probability of observing one's
  data, given the assumed model. The presense of selection effects and
  intrinsic uncertainty in the broad line mass estimates can make this
  difficult, as the observed data likelihood function is not simply
  given by Equation (\ref{eq-phiconvert}). However, we can account for
  these difficulties by first deriving the likelihood function for the
  complete set of data, and then integrating over the missing data to
  obtain the observed data likelihood function.

  For broad line AGNs, we can relate the distribution of $M_{BH}$ and
  $z$ to the joint distribution of $L_{\lambda}, {\bf v}$, and
  $z$. Here, ${\bf v} = (v_{H\beta}, v_{MgII}, v_{CIV})$, where
  $v_{H\beta} = v_{H\beta}$ is the the velocity dispersion for the
  H$\beta$ broad line emitting gas, and similarly for $v_{MgII}$ and
  $v_{CIV}$. These three lines are commonly used in estimating
  $M_{BH}$ from single-epoch spectra of broad line AGN
  \citep[e.g.,][]{mclure02,kaspi05,vest02,vest06}, where the velocity
  dispersion is typically estimated from the $FWHM$ of the emission
  line. The distribution of $L_{\lambda}$ and ${\bf v}$ are then
  related to the BHMF via the $R$--$L$ relationship and the virial
  theorem.

  The BHMF for broad line AGN can be inferred from the distribution of
  $L_{\lambda}, {\bf v},$ and $z$, and thus it is necessary to
  formulate the observed data likelihood function in terms of
  $(L_{\lambda}, {\bf v}, z)$. While it is possible to formulate the
  likelihood function in terms of the broad line mass estimates,
  denoted as $\hat{M}_{BL} \propto L_{\lambda}^{1/2} V^2$, the
  logarithm of the broad line mass estimates are simply linear
  combinations of $\log L_{\lambda}$ and $\log {\bf v}$, and thus
  statistical inference does not depend on whether we formulate the
  likelihood function in terms of $L_{\lambda}$ and ${\bf v}$ or
  $\hat{M}_{BL}$. We find it mathematically simpler and more intuitive
  to infer the BHMF directly from the distribution of $L_{\lambda},
  {\bf v},$ and $z$, as opposed to inferring it from the distribution
  of $L_{\lambda}, \hat{M}_{BL},$ and $z$.

  Following the discussion in KFV08, we derive the likelihood function
  for the set of observed luminosities, redshifts, and emission line
  widths. We introduce an indicator variable $I$ denoting whether a
  source is included in the survey or not: if $I_i=1$ then a source is
  included, otherwise, $I_i = 0$. The variable $I$ is considered to be
  part of the observed data in the sense that we `observe' whether a
  source is detected or not. The survey selection function is the
  probability of including the $i^{\rm th}$ source in one's survey,
  $p(I_i = 1|{\bf v}_i,L_{\lambda,i},z_i)$. Here, we have assumed that
  the probability of including a source in one's sample only depends
  on luminosity, redshift, and emission line width, and is therefore
  conditionally independent of $M_{BH}$. This is the case, in general,
  since one can only select a survey based on quantities that are
  directly observable. Including the additional `data' $I$, the
  observed data likelihood function for broad line AGN is:
  \begin{eqnarray}
    \lefteqn{p({\bf v}_{obs},L_{obs},z_{obs},I|\theta,N) \propto} \\
    & & C^N_n \prod_{i \in {\cal A}_{obs}} \int p({\bf v}_i,L_{\lambda,i},M_{BH,i},z_i|\theta) \ dM_{BH,i} \\
    & & \times \prod_{j \in {\cal A}_{mis}} \int \int \int \int p(I=0|{\bf v}_j, L_{\lambda,j}, z_j) 
    p({\bf v}_j,L_{\lambda,j},M_{BH,j},z_j|\theta)\ d{\bf v}_j\ dL_{\lambda,j} dM_{BH,j}\ dz_j \\
    & \propto & C^N_n \left[ p(I=0|\theta) \right]^{N-n} 
    \prod_{i \in {\cal A}_{obs}} p({\bf v}_i,L_{\lambda,i},z_i|\theta),
    \label{eq-blobslik}
  \end{eqnarray}
  where ${\cal A}_{obs}$ denotes the set of sources included in one's
  survey, ${\cal A}_{mis}$ denotes the set of sources not included in
  one's survey, and on the last line we have omitted terms that do not
  depend on $N$ or $\theta$. Here,
  \begin{equation}
    p({\bf v}_i,L_{\lambda,i},z_i|\theta) = \int_{0}^{\infty} 
    p({\bf v}_i,L_{\lambda,i},z_i,M_{BH,i}|\theta)\ dM_{BH,i}
    \label{eq-likvlz}
  \end{equation}
  is the probability of observing values of ${\bf v}_i,L_{\lambda,i}$, and $z_i$
  for the $i^{\rm th}$ source, given $\theta$, and
  \begin{equation}
    p(I=0|\theta) = \int_{0}^{\infty} 
    \int_{0}^{\infty} \int_{0}^{\infty} p(I=0|{\bf v}, L_{\lambda}, z) 
    p({\bf v},L_{\lambda},z|\theta)\ d{\bf v}\ dL_{\lambda}\ dz
    \label{eq-misprob}
  \end{equation}
  is the probability that the survey misses a source, given $\theta$;
  note that $p(I=0|\theta) = 1 - p(I=1|\theta)$. Qualitatively, the
  observed data likelihood function for the BHMF is the probability of
  observing a set of $n$ emission line widths ${\bf v}_1,\ldots,{\bf
  v}_n$, luminosities $L_{\lambda,1}, \ldots, L_{\lambda,n}$, and
  redshifts $z_1, \ldots, z_n$ given the assumed BHMF model
  parameterized by $\theta$, multiplied by the probability of not
  detecting $N - n$ sources given $\theta$, multiplied by the number
  of ways to select a subset of $n$ sources from a set of $N$ total
  sources. Equation (\ref{eq-blobslik}) can be maximized to calculate
  a maximum likelihood estimate of the black hole mass function when
  using broad line estimates of $M_{BH}$, or combined with a prior
  distribution to perform Bayesian inference.

  It is often preferred to write the BHMF observed data likelihood
  function by factoring the joint distribution of ${\bf v},
  L_{\lambda}, M_{BH},$ and $z$ into conditional distributions. This
  has the advantage of being easier to interpret and work with,
  especially when attempting to connect the distribution of line
  widths and luminosities to the distribution of black hole mass. The
  joint distribution can be factored as \citep{kelly07b}
  \begin{equation}
    p({\bf v},L_{\lambda},M_{BH},z) = p({\bf v}|L_{\lambda},M_{BH},z) p(L_{\lambda}|M_{BH},z) 
    p(M_{BH},z).
    \label{eq-jointdist}
  \end{equation}
  Here, $p({\bf v}|L_{\lambda},M_{BH},z)$ is the distribution of
  emission line widths at a given $L_{\lambda}, M_{BH},$ and $z$,
  $p(L_{\lambda}|M_{BH},z)$ is the distribution of luminosities at a
  given $M_{BH}$ and $z$, and $p(M_{BH},z)$ is the probability
  distribution of black hole mass and redshift, related to the BHMF
  via Equation (\ref{eq-phiconvert}). When using broad line estimates
  of $M_{BH}$, it is assumed that $p({\bf v}|L_{\lambda},M_{BH},z)$ is
  set by the virial theorem, where the distance between the central
  black hole and the broad line-emitting gas depends on $L_{\lambda}$
  via the $R$--$L$ relationship. In this work we assume that the
  $R$--$L$ relationship does not depend on $z$
  \citep[e.g.,][]{vest04}, and thus $p({\bf v}|L_{\lambda},M_{BH},z) =
  p({\bf v}|L_{\lambda},M_{BH})$.

  Under the factorization given by Equation (\ref{eq-jointdist}), the
  observed data likelihood function (Eq. [\ref{eq-blobslik}]) becomes
  \begin{eqnarray}
    \lefteqn{p({\bf v}_{obs},L_{obs},z_{obs},I|\theta,N) \propto} \nonumber \\
    & & C^N_n \left[ p(I=0|\theta) \right]^{N-n} 
    \prod_{i \in {\cal A}_{obs}} \int_{0}^{\infty} p({\bf v}_i|L_{\lambda,i},M_{BH,i},\theta) 
    p(L_{\lambda,i}|M_{BH,i},z,\theta) p(M_{BH,i},z|\theta)\ dM_{BH,i}.
    \label{eq-blobslik2}
  \end{eqnarray}
  The BHMF likelihood function, given by Equation (\ref{eq-blobslik})
  or (\ref{eq-blobslik2}), is entirely general, and it is necessary to
  assume parametric forms in order to make use of it. In
  \S~\ref{s-smodel} we describe a parametric form based on a mixture
  of Gaussian functions model, and explicitly calculate Equation
  (\ref{eq-blobslik}) for the mixture model.

  \subsection{Selection Function}

  \label{s-selfunc}

  The selection probability, $p(I=1|{\bf v},L_{\lambda},z)$, depends
  on both the luminosity and redshift through the usual flux
  dependence, but can also depend on the emission line width.  In
  particular, an upper limit on ${\bf v}$ may occur if there is a
  width above which emission lines become difficult to distinguish
  from the continuum and iron emission.  In this case, if all emission
  lines in one's spectrum are wider than the maximum line width than
  one is not able to obtain a reliable estimate of the line width for
  any emission line, and therefore the source is not used to estimate
  $\phi(M_{BH},z)$. A lower limit on the line width may be imposed in
  order to prevent the inclusion of narrow line AGN, for which broad
  line mass estimates are not valid. In this case the inclusion
  criterion might be that at least one emission line is broader than,
  say, $FWHM = 2000 {\rm\ km\ s^{-1}}$.  In addition to the limits on
  line width that may be imposed, there is an upper and lower limit on
  $z$ due to redshifting of emission lines out of the observable
  spectral range. For example, if one uses optical spectra than the
  range of useable spectra is $0 < z \lesssim 4.5$, as the C IV line
  redshifts into the near-infrared for $z \gtrsim 4.5$.

  Denote the upper and lower limit of ${\bf v}$ as $v_{min}$ and
  $v_{max}$, and the upper and lower limit of $z$ as $z_{min}$ and
  $z_{max}$. Furthermore, denote the usual survey selection function
  in terms of $L_{\lambda}$ and $z$ as $s(L_{\lambda},z)$, where
  $s(L_{\lambda},z)$ is the probability that a source is included in
  the survey before any cuts on line width are imposed;
  $s(L_{\lambda},z)$ would typically correspond to the selection
  function used in luminosity function estimation. Note that in this
  work $s(L_{\lambda},z)$ gives the probability that any source in the
  universe is included in the survey, given its luminosity and
  redshift, and thus $s(L_{\lambda},z) \leq \Omega / 4\pi$, where
  $\Omega / 4\pi$ is the fraction of the sky covered by the
  survey. Then, $p(I_i=1|{\bf v}_i,L_{\lambda,i},z_i) =
  s(L_{\lambda,i},z_i)$ if $z_{min} \leq z_i \leq z_{max}$ and at
  least one emission line has $v_{min} \leq v_i \leq v_{max}$;
  otherwise, $p(I_i=1|{\bf v}_i,L_{\lambda,i},z_i) = 0$. In this case,
  the probability that a source is included in the survey (see
  Eq.[\ref{eq-misprob}]) is
  \begin{eqnarray}
    \lefteqn{p(I=1|\theta) = } \nonumber \\
    & & \int_{0}^{\infty} \int_{z_{min}}^{z_{max}} s(L_{\lambda},z) \int_{v_{min}}^{v_{max}} 
      \int_{0}^{\infty} p({\bf v}|L_{\lambda},M_{BH},\theta) p(L_{\lambda}|M_{BH},z,\theta) 
      p(M_{BH},z|\theta)\ dM_{BH}\ d{\bf v}\ dz\ dL_{\lambda},
    \label{eq-selprob}
  \end{eqnarray}
  where the inner two integrals are over ${\bf v}$ and $M_{BH}$, and
  the outer two integrals are over $L_{\lambda}$ and $z$. One can then
  insert Equation (\ref{eq-selprob}) into Equation (\ref{eq-blobslik})
  to get the likelihood function.

  It is informative to express the selection function in terms of
  black hole mass and redshift. The selection function as a function
  of black hole mass and redshift is the probability of including a
  source, given its $M_{BH}$ and $z$, and is calculated as
  \begin{equation}
    p(I = 1|M_{BH},z) =  \int_{0}^{\infty} s(L_{\lambda},z) p(L_{\lambda}|M_{BH},z) 
    \int_{v_{min}}^{v_{max}} p({\bf v}|L_{\lambda},M_{BH})\ dL_{\lambda}\ d{\bf v}.
    \label{eq-mselect_convert}
  \end{equation}
  At any given value of $M_{BH}$ a range of luminosities and emission
  line widths are possible, and thus sources with low black hole mass
  can be detected if they are bright enough and have line widths
  $v_{min} < v < v_{max}$. Conversely, sources with high black hole
  masses can be missed by the survey if their luminosity is below the
  flux limit at that redshift, or if their line width falls outside of
  the detectable range. This has the effect of smoothing the survey's
  selection function, and thus the black hole mass selection function
  is a broadened form of the flux selection function.

  As an example, consider the case when the selection function is
  simply a flux limit. In this case, the selection function is
  \begin{equation}
    s(l,z) = \left\{ \begin{array}{ll}
      1\ {\rm if}\ 4\pi f_{min} D^2_L(z) < L_{\lambda} < 4\pi f_{max} D^2_L(z) \\
      0\ {\rm otherwise} \end{array} \right.,
    \label{eq-sfunc_example}
  \end{equation}
  where $f_{min}$ is the survey's lower flux limit, $f_{max}$ is the
  survey's upper flux limit, and $D_L(z)$ is the luminosity distance
  to redshift $z$. For simplicity, in this example we assume that
  there is no additional cut on emission line width, i.e., $v_{min} =
  0$ and $v_{max} = \infty$. In this case, the black hole mass
  selection function, $p(I=1|M_{BH},z)$, is the convolution of the
  luminosity selection function with the distribution of $L_{\lambda}$
  at a given $M_{BH}$. If the distribution of $\log L_{\lambda}$ at a
  given $M_{BH}$ is a Gaussian function with mean $\alpha_0 + \alpha_m
  \log M_{BH}$ and dispersion $\sigma_l$, then the black hole mass
  selection function is
  \begin{equation}
    p(I=1|M_{BH},z) = 
    \Phi\left(\frac{\log L_{max}(z) - \alpha_0 - \alpha_m \log M_{BH}}{\sigma_l} \right) - 
    \Phi\left(\frac{\log L_{min}(z) - \alpha_0 - \alpha_m \log M_{BH}}{\sigma_l} \right).
    \label{eq-msfunc_example}
  \end{equation}
  Here, $L_{max}(z) = 4\pi f_{max} D^2_L(z)$, $L_{min}(z) = 4\pi
  f_{min} D^2_L(z)$, and $\Phi(\cdot)$ is the cumulative distribution
  function of the standard normal distribution.

\begin{figure}
    \begin{center}
      \scalebox{0.5}{\rotatebox{90}{\plotone{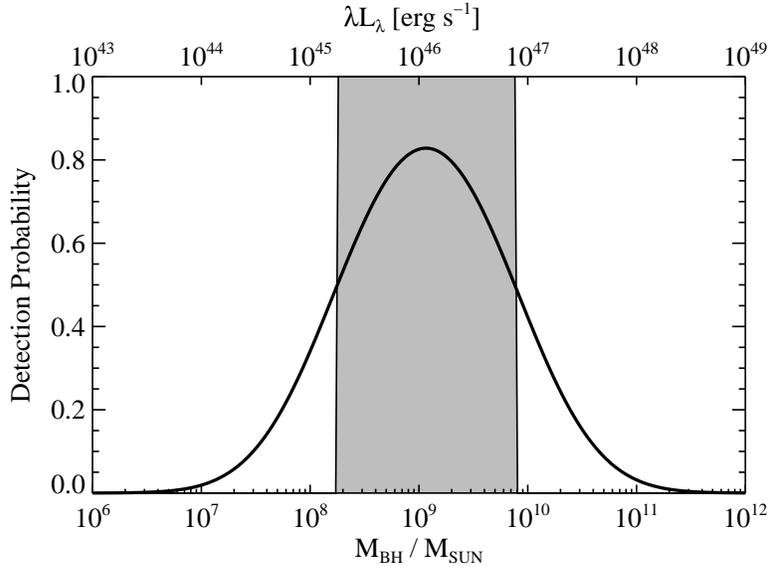}}}
      \caption{Comparison of the selection function for luminosity
        (shaded region) and black hole mass (curve) for a simple upper and
        lower flux limit. The selection is complete in luminosity within
        the flux limits, but is never `complete' in $M_{BH}$. The
        intrinsic physical range in luminosity at a given black hole
        mass creates a more complicated selection function for $M_{BH}$,
        since at any given $M_{BH}$ and $z$ only those quasars with
        luminosities within the flux limits are detected.
        \label{f-msfunc_example}}
    \end{center}
  \end{figure}

  In Figure \ref{f-msfunc_example} we show the black hole mass
  selection function, $p(I=1|M_{BH},z)$, given by Equation
  (\ref{eq-msfunc_example}) at $z = 1$. Here, we have used the SDSS
  quasar sample flux limit, $19.1 > i > 15$, $\alpha_0 = 37, \alpha_m
  = 1,$ and $\sigma_l = 0.6$ dex. Because the black hole mass
  selection function is the convolution of the luminosity selection
  function with the distribution of $L_{\lambda}$ at a given $M_{BH}$,
  the black hole mass selection function is positive over a wider
  range in $M_{BH}$, as compared to the range in $L_{\lambda}$ for
  which $s(L_{\lambda},z)$ is positive. However, because
  $p(I=1|M_{BH},z)$ spreads the selection probability over a wider
  range in $M_{BH}$, all bins in $M_{BH}$ are incomplete.

  The difference in selection functions for black hole mass and
  luminosity results in an important distinction between the
  estimation of black hole mass functions and the estimation of
  luminosity functions. First, one cannot correct the binned BHMF for
  the survey flux limits by simply applying the $1 / V_a$
  correction. This is a common technique used for estimating binned
  luminosity functions, where the number density in a
  $(L_{\lambda},z)$ bin is corrected using the survey volume in which
  a source with luminosity $L_{\lambda}$ could have been detected and
  still remained in the redshift bin. In the case of the BHMF, a
  survey volume in which the black hole could have been detected
  ceases to have any meaning, as black holes can be detected over many
  different survey volumes, albeit with varying
  probability. Alternatively, the $1 / V_a$ correction can be thought
  of as dividing the number of sources in a bin in $(L_{\lambda},z)$
  by the detection probability as a function of $L_{\lambda}$ and
  $z$. Therefore, applying a $1 / V_a$ correction to a bin in
  $(M_{BH},z)$ is essentially the same as dividing the number of
  sources in a bin in $(M_{BH},z)$ by the detection probability as a
  function of $L_{\lambda}$ and $z$. For the simple example shown in
  Figure \ref{f-msfunc_example}, those quasars in a given bin in
  $(M_{BH},z)$ that happen to have luminosities $L_{min}(z) <
  L_{\lambda} < L_{max}(z)$ will receive no correction, since
  $s(L_{\lambda},z) = 1$. However, those quasars which have
  luminosities outside of the detectable range will not be
  detected. The end result is a systematic underestimate of the binned
  BHMF.

  The number of sources in a given bin in the $M_{BH}$--$z$ plane can
  be estimated by dividing the observed number of black holes in each
  bin by the black hole mass selection function,
  $p(I=1|M_{BH},z)$. Similarly, one can use a $1 / V_a$-type
  correction by calculating an `effective' $1 / V_a$, found by
  integrating $dV / dz$ over the black hole mass selection
  function. This approach has been adopted previously within the
  context of binned luminosity functions
  \citep[e.g.,][]{warren94,fan01}. However, it is essential that the
  black hole mass selection function be used and \emph{not} the
  luminosity selection function. Unfortunately, this implies that one
  must assume a form for $p(L_{\lambda}|M_{BH},z)$.

  \section{POSTERIOR DISTRIBUTION FOR THE BHMF PARAMETERS}

  \label{s-posterior}

  The posterior probability distribution of the model parameters is
  \begin{equation}
    p(\theta, N|{\bf v}_{obs},L_{obs},z_{obs},I) \propto p(\theta, N)
    p({\bf v}_{obs},L_{obs},z_{obs},I | \theta, N),
    \label{eq-post}
  \end{equation}
  where $p(\theta, N)$ is the prior on $(\theta, N)$, and $p({\bf
  v}_{obs}, L_{obs}, z_{obs}, I | \theta, N)$ is the likelihood
  function, given by Equation (\ref{eq-blobslik}). The posterior
  distribution gives the probability that $\theta$ and $N$ have a
  given value, given the observed data $({\bf v}_{obs}, L_{obs},
  z_{obs})$. Therefore, the posterior distribution of $\theta$ and $N$
  can be used to obtain the probability that $\phi(M_{BH},z)$ has any
  given value, given that we have observed some set of emission line
  widths, luminosities, and redshifts.

  It is of use to decompose the posterior as $p(N,\theta|x_{obs})
  \propto p(N|\theta,x_{obs}) p(\theta|x_{obs})$, where we have
  abbreviated the observed data as $x_{obs} = ({\bf v}_{obs}, L_{obs},
  z_{obs})$. This decomposition seperates the posterior into the
  conditional posterior of the BHMF normalization,
  $p(N|x_{obs},\theta)$, from the marginal posterior of the BHMF
  shape, $p(\theta|x_{obs})$. In this work we take $N$ and $\theta$ to
  be independent in their prior distribution, $p(N,\theta) = p(N)
  p(\theta)$, and that the prior on $N$ is uniform over $\log N$. In
  this case, one case show \citep[e.g.,][KFV08]{gelman04} that the
  marginal posterior distribution of $\theta$ is
  \begin{equation}
    p(\theta|{\bf v}_{obs},L_{obs},z_{obs}) \propto p(\theta) 
    \left[p(I=1|\theta) \right]^{-n} \prod_{i \in {\cal A}_{obs}} 
    p({\bf v}_i,L_{\lambda,i},z_i|\theta),
    \label{eq-thetapost}
  \end{equation}
  where $p(I=1|\theta) = 1 - p(I=0|\theta)$. 

  Under the prior $p(\log N) \propto 1$, the conditional posterior of
  $N|\theta,x_{obs}$ is a negative binomial distribution with
  parameters $n$ and $p(I=1|\theta)$. The negative binomial
  distribution gives the probability that the total number of sources
  is equal to $N$, given that there have been $n$ detections with
  probability of detection $p(I=1|\theta)$:
  \begin{equation}
    p(N|n,\theta) = C^{N-1}_{n-1} \left[p(I=1|\theta)\right]^n \left[p(I=0|\theta)\right]^{N-n}.
    \label{eq-npost}
  \end{equation}
  
  Because of the large number of parameters in the model, Bayesian
  inference is most easily performed by randomly drawing values of $N$
  and $\theta$ from their posterior. Based on the decomposition
  $p(\theta,N|x_{obs}) \propto p(N|n,\theta) p(\theta|x_{obs})$, we
  can obtain random draws of $(N,\theta)$ by first drawing values of
  $\theta$ from Equation (\ref{eq-thetapost}). Then, for each draw of
  $\theta$, we draw a value of $N$ from the negative binomial
  distribution. Random draws for $\theta$ may be obtained via markov
  chain monte carlo (MCMC) methods, described in \S~\ref{s-mha}, and
  random draws from the negative binomial distribution are easily
  obtained using standard methods \citep[e.g.,][KFV08]{gelman04}.

  \section{THE STATISTICAL MODEL}

  \label{s-smodel}

  In order to compute the likelihood function for the observed set of
  luminosities, redshifts, and broad emission line widths (see
  Eq.[\ref{eq-blobslik2}]), it is necessary to relate the BHMF to the
  distribution of $L_{\lambda}$ and ${\bf v}$. To do this, Equation
  (\ref{eq-jointdist}) implies that we need three terms. The first
  term is an assumed BHMF, $p(M_{BH},z) = N^{-1}(dV / dz)^{-1}
  \phi(M_{BH},z)$. The second term is an assumed distribution of
  luminosities at a given black hole mass and redshift,
  $p(L_{\lambda}|M_{BH},z)$. The third term is an assumed distribution
  of broad emission line widths at a given luminosity and black hole
  mass, $p({\bf v}|L_{\lambda},M_{BH})$. Once we have a parameteric
  form for each of these three distributions, we can calculate the
  observed data likelihood directly from Equation
  (\ref{eq-blobslik2}). In this section we describe parameteric forms
  for each of these distributions based on a mixture of Gaussian
  functions model.

  \subsection{Mixture of Gaussian Functions Model for the BHMF}

  \label{s-probmz}

  The mixture of Gaussian functions model is a common
  `non-parameteric' model that allows flexibility when estimating the
  BHMF. The basic idea is that one can use a suitably large enough
  number of Gaussian functions to accurately approximate the true
  BHMF, even though the individual Gaussian functions have no physical
  meaning. Furthermore, the Gaussian mixture model is also conjugate
  to the distributions $p(L_{\lambda}|m)$ and $p({\bf
  v}|L_{\lambda},m)$ assumed in \S\S \ref{s-problm} and \ref{s-vprob},
  thus enabling us to calculate some of the integrals in Equation
  (\ref{eq-thetapost}) analytically.

  In KFV08 we described a mixture of Gaussian functions model for a
  luminosity function. The mixture of Gaussian functions model of the
  BHMF is identical to that for the luminosity function, after
  replacing $L$ with $M_{BH}$. Our mixture of Guassian functions
  model, including our adopted prior, is described in KFV08; for
  completeness we briefly review it here.

  The mixture of $K$ Gaussian functions model for the BHMF is
  \begin{equation}
    p(\log M_{BH},\log z|\pi, \mu, \Sigma) = \sum_{k=1}^{K} \frac{\pi_k}{2 \pi
      |\Sigma_k|^{1/2}} \exp \left[ -\frac{1}{2} ({\bf y} - \mu_k)^T
      \Sigma_k^{-1} ({\bf y} - \mu_k) \right],
    \label{eq-mixmod}
  \end{equation}
  where $\sum_{k=1}^{K} \pi_k = 1$. Here, ${\bf y} = (\log M_{BH},\log
  z)$, $\mu_k$ is the 2-element mean vector for the $k^{\rm th}$
  Gaussian functions, $\Sigma_k$ is the $2 \times 2$ covariance matrix for the
  $k^{\rm th}$ Gaussian function, and $x^T$ denotes the transpose of $x$. In
  addition, we denote $\pi = (\pi_1, \ldots, \pi_K), \mu = (\mu_1,
  \ldots, \mu_K)$, and $\Sigma = (\Sigma_1, \ldots, \Sigma_K)$. The
  variance in $\log M_{BH}$ for Gaussian function $k$ is $\sigma^2_{m,k} =
  \Sigma_{11,k}$, the variance in $\log z$ for Gaussian function $k$ is
  $\sigma^2_{z,k} = \Sigma_{22,k}$, and the covariance between $\log
  M_{BH}$ and $\log z$ for Gaussian function $k$ is $\sigma_{mz,k} =
  \Sigma_{12,k}$. Note that Equation(\ref{eq-mixmod}) is equivalent to
  assuming that $p(M_{BH},z)$ is a mixture of log-normal
  densities. Under the mixture model, the BHMF can be calculated from
  Equations (\ref{eq-phiconvert}) and (\ref{eq-mixmod}). Noting that
  $p(M_{BH},z) = p(\log M_{BH},\log z) / (M_{BH} z (\ln 10)^2)$, the
  mixture of normals model for the BHMF is
  \begin{equation}
    \phi(M_{BH},z|\theta,N) = \frac{N}{M_{BH} z (\ln 10)^2} \left( \frac{dV}{dz} \right)^{-1}
    \sum_{k=1}^{K} \frac{\pi_k}{2 \pi |\Sigma_k|^{1/2}} \exp 
    \left[ -\frac{1}{2} ({\bf y} - \mu_k)^T \Sigma_k^{-1} ({\bf y} - \mu_k) \right] 
    \label{eq-mixbhmf},
  \end{equation}
  where, as before, ${\bf y} = (\log M_{BH}, \log z)$.

  \subsection{The Distribution of $L_{\lambda}$ at a Given $M_{BH}$}

  \label{s-problm}
  
  We model the distribution of luminosities at a given $M_{BH}$ as a
  log-normal distribution, where the average $\log L_{\lambda}$ at a given
  $M_{BH}$ depends linearly on $\log M_{BH}$:
  \begin{equation}
    p(\log L_{\lambda}|M_{BH},\alpha) = \frac{1}{\sqrt{2 \pi \sigma_l^2}} \exp \left[
      -\frac{1}{2} \left( \frac{\log L_{\lambda} - \alpha_0 - \alpha_m \log M_{BH}}{\sigma_l}
      \right)^2 \right].
    \label{eq-problm}
  \end{equation}
  Here, the unknown parameters are $\alpha = (\alpha_0, \alpha_m,
  \sigma_l^2)$. This is equivalent to assuming a simple linear
  regression of $\log L_{\lambda}$ on $\log M_{BH}$, where $\alpha_0$
  is the constant, $\alpha_m$ is the slope, and $\sigma_l$ is the
  standard deviation of the random Gaussian dispersion about the
  regression line. We assume a uniform prior on these parameters,
  i.e., $p(\alpha_0,\alpha_m, \sigma_l) \propto 1$.

  The form of the $M_{BH}$--$L_{\lambda}$ relationship given by
  Equation (\ref{eq-problm}) is motivated by noting that $L_{\lambda}$
  can be related to $M_{BH}$ as
  \begin{equation}
    \lambda L_{\lambda} = 1.3 \times 10^{38} \frac{\Gamma_{Edd}}{C_{\lambda}}
    \frac{M_{BH}}{M_{\odot}} \ \ [{\rm erg\ s^{-1}}],
      \label{eq-mlrel}
  \end{equation}
  where $\Gamma_{Edd} \equiv L_{bol} / L_{Edd}$ is the Eddington
  ratio, and $C_{\lambda}$ is the bolometric correction to $\lambda
  L_{\lambda}$. Equation (\ref{eq-mlrel}) implies that the
  distribution of luminosities at a given black hole mass is caused by
  the distribution in Eddington ratios and bolometric corrections at a
  given black hole mass. The distribution of $\log L_{\lambda}$ at a
  given $M_{BH}$ is the convolution of the distribution of $\log
  \Gamma_{Edd}$ at a given $M_{BH}$, with the distribution of $\log
  C_{\lambda}$ at a given $M_{BH}$. The parameter $\sigma_l$ is thus
  an estimate of the dispersion in $\log (\Gamma_{Edd} / C_{\lambda})$
  at a given $M_{BH}$.

  If both $\Gamma_{Edd}$ and $C_{\lambda}$ are statistically
  independent of $M_{BH}$, then we would expect that on average
  $L_{\lambda} \propto M_{BH}$, i.e., $\alpha_m = 1$. However, if
  $\Gamma_{Edd}$ or $C_{\lambda}$ are correlated with $M_{BH}$, then
  $\alpha_m \neq 1$. Currently, it is unknown whether $M_{BH}$ and
  $\Gamma_{Edd}$ are correlated. However, it is likely that quasar
  SEDs depend on both $\Gamma_{Edd}$ and $M_{BH}$, and therefore the
  bolometric correction will also depend on $\Gamma_{Edd}$ and
  $M_{BH}$. Indeed, recently some authors have found evidence that the
  bolometric correction depends on Eddington ratio \citep{vasud07} and
  black hole mass \citep{kelly08}. Therefore, it is
  likely that $\alpha_m \neq 1$, and we leave it as a free
  parameter. In addition, comparison of Equation (\ref{eq-problm})
  with Equation (\ref{eq-mlrel}), and assuming that $\Gamma_{Edd} /
  C_{\lambda}$ is independent of $M_{BH}$, implies that the average
  value of $\Gamma_{Edd} / C_{\lambda}$ is related to $\alpha_0$
  according to $E(\log \Gamma_{Edd} / C_{\lambda}) = \alpha_0 -
  38.11$, where $E(x)$ denotes the expectation value of
  $x$. Therefore, one can use $\alpha_0$ to estimate the typical broad
  line quasar Eddington ratio, assuming a typical bolometric
  correction.

  Currently, there is little known about the distribution of
  luminosities at a given black hole mass, so for simplicity we assume
  the simple linear form given by Equation
  (\ref{eq-problm}). Furthermore, the assumption of a Gaussian
  distribution in $\log L$ at a given $M_{BH}$ is consistent with the
  $L$--$M_{BH}$ relationship for those AGN with reverberation mapping
  data \citep{kelly07b}. More sophisticated models could include a
  non-linear dependence on $\log M_{BH}$, an additional redshift
  dependence, or non-Gaussian distribution. Unfortunately, this
  introduces additional complexity into the model. Furthermore, an
  additional redshift dependence in Equation (\ref{eq-problm}) implies
  that the distribution of $\Gamma_{Edd}$ or $C_{\lambda}$ at a given
  $M_{BH}$ evolves. However, currently most investigations have not
  found any evidence for significant evolution in $\Gamma_{Edd}$
  \citep[e.g.,][]{corbett03,mclure04,vest04,koll06}, and it is unclear
  if the quasar SED evolves at a given $M_{BH}$. Therefore, there is
  currently no compelling evidence to justify inclusion of a redshift
  dependence in Equation (\ref{eq-problm}). In addition, we note that
  it is impossible to use $p(L|M_{BH})$ to infer the distribution of
  Eddington ratios without making an assumption about the distribution
  of $C_{bol}$, as Equation (\ref{eq-mlrel}) shows that $\Gamma_{Edd}$
  and $C_{bol}$ are degenerate. While estimating the distribution of
  $\Gamma_{Edd}$ is of significant interest, it is beyond the scope of
  this work to develop a robust technique to do so, as our goal is to
  estimate the black hole mass function.

  Because of the large number of parameters, large uncertainty in the
  broad line black hole mass estimates, and flux limit, estimating the
  BHMF is already a difficult statistical problem. As such, our
  approach is to initially assume the simple form given by Equation
  (\ref{eq-problm}) in order to keep the degrees of freedom low, and
  to check if this assumption is consistent with our data (see
  \S~\ref{s-postcheck}). If it is found that the observed data are
  inconsistent with this statistical model (e.g., see
  \S~\ref{s-postcheck}) then Equation (\ref{eq-problm}) should be
  modified.

  \subsection{The Distribution of $v$ at a given $L$ and $M_{BH}$}

  \label{s-vprob}

  Following \citet{kelly07b}, we can derive the distribution of
  emission line widths at a given luminosity and black hole
  mass. Given an AGN luminosity, $L_{\lambda}^{BL}$, the BLR distance
  $R$ is assumed to be set by the luminosity according to the $R$--$L$
  relationship, $R \propto L_{\lambda}^{\beta_l}$, with some
  additional log-normal statistical scatter:
  \begin{equation}
    p(\log R|L^{BL}_{\lambda}) = \frac{1}{\sqrt{2 \pi \sigma_r^2}} \exp \left[
      -\frac{1}{2} \left( \frac{\log R - r_0 - \beta_l \log L^{BL}_{\lambda}}{\sigma_r}
      \right)^2 \right].
    \label{eq-rlum}
  \end{equation}
  Here, $r_0$ is a constant, $\sigma_r$ is the dispersion in $\log R$
  at a given luminosity, and $L_{\lambda}^{BL}$ is the AGN continuum
  luminosity at some reference wavelength appropriate for the broad
  emission line of interest. Note that the reference wavelength for
  $L^{BL}_{\lambda}$ is not necessarily the same wavelength as for
  $L_{\lambda}$ used in \S~\ref{s-problm}. In particular, the
  wavelength for $L_{\lambda}$ used in the $M_{BH}$--$L_{\lambda}$
  relationship should be chosen to adequately account for the
  selection function, while the reference wavelength for
  $L^{BL}_{\lambda}$ should be appropriate for describing the $R$--$L$
  relationship. Since AGN continua are well described by a power-law,
  $f_{\nu} \propto \nu^{-\alpha}$, it should be easy to calculate
  $L_{\lambda}$ at different values so long as the spectral index,
  $\alpha$, is known. The intrinsic scatter in $R$ at a given
  $L^{BL}_{\lambda}$ is likely due to variations in quasar SED,
  reddening, non-instantaneous response of the BLR to continuum
  variations, etc.

  Assuming that the BLR gas is gravitationally bound, the velocity
  dispersion of the broad line-emitting gas is related to $R$ and
  $M_{BH}$ as $M_{BH} = f R v^2 / G $. Here, $G$ is the gravitational
  constant, and $f$ is a factor that converts the virial product, $R
  M_{BH} / G$, to a mass. We do not directly measure $v$, but instead
  estimate it by the $FWHM$ or dispersion of the broad emission line
  in a single-epoch spectra. As a result, the measured line width will
  scatter about the actual value of $v$, where this scatter may be due
  in part to variations in line profile shape and the existence of
  stationary components in the single-epoch line profile. In our
  statistical model we assume that this scatter is log-normal with a
  dispersion of $\sigma_v$. In addition, the value of $f$ depends on
  the measure of line width used. \citet{onken04} estimated $f$ by
  comparing black hole masses derived form reverberation mapping with
  those derived from the $M_{BH}$--$\sigma$ relationship, and find
  that on average $f = 1.4 \pm 0.4$ when using the $FWHM$. This value
  is consistent with a value of $f = 0.75$ expected from a spherical
  BLR geometry \citep[e.g.,][]{netzer90}.

  Under our model, the distribution of emission line widths at a given
  BLR size and black hole mass is
  \begin{equation}
    \log v|R,M_{BH} = \frac{1}{\sqrt{2 \pi \sigma_v^2}} \exp \left\{
      -\frac{1}{2} \left[ \frac{\log v - v_0 - 1/2 (\log f + \log R - 
	\log M_{BH})}{\sigma_v}\right]^2 \right\}.
    \label{eq-vrm}
  \end{equation}
  where $v_0$ is a constant. For convenience, here and throughout this
  paper we denote the estimate of the BLR gas velocity dispersion as
  $v$, i.e., $v$ is either the $FWHM$ or dispersion of the emission
  line. The term $v$ in Equation (\ref{eq-vrm}) should not be confused
  with the actual velocity dispersion of the BLR gas, but is an
  estimate of it based on a measure of the width of the broad emission
  line. From Equation (\ref{eq-vrm}) it is apparent that the term $f$
  shifts the distribution of $\log v$ by a constant amount, which has
  the effect of shifting the inferred BHMF by a constant amount in
  $\log M_{BH}$. Throughout the rest of this work we assume the value
  of $f = 1.4$ found by \citet{onken04}.

  The distribution of $v$ at a given $L$ and $M_{BH}$ is obtained from
  Equations (\ref{eq-rlum}) and (\ref{eq-vrm}) by averaging the
  distribution of $v$ at a given $R$ and $M_{BH}$ over the
  distribution of $R$ at a given $L^{BL}_{\lambda}$:
  \begin{eqnarray}
    p(\log v|L^{BL}_{\lambda},M_{BH},\beta) & = & \int_{-\infty}^{\infty} 
    p(\log v|\log R,M_{BH},\beta) p(\log R|L_{\lambda}^{BL},\beta)\ d\log R
    \label{eq-vlm0} \\
    & = & \frac{1}{\sqrt{2 \pi \sigma_{BL}^2}} \exp \left\{
      -\frac{1}{2} \left[ \frac{\log v - \beta_0 - 1/2 (\beta_l \log L_{\lambda}^{BL} - 
	\log M_{BH})}{\sigma_{BL}}\right]^2 \right\}.
    \label{eq-probvlm}
  \end{eqnarray}
  where $\beta_0$ is a constant, $\sigma^2_{BL} = \sigma^2_v +
  \sigma^2_r / 4$, and $\beta \equiv (\beta_0, \beta_l,
  \sigma_{BL})$. Note that in Equation (\ref{eq-probvlm}) we have
  absorbed $\log f$ into the constant term, $\beta_0$. The term
  $\sigma_{BL}$ is the dispersion in emission line widths at a given
  luminosity and black hole mass, and can be related to the intrinsic
  uncertainty in the broad line estimates of $M_{BH}$. The usual broad
  line mass estimates of AGN can be obtained by reexpressing the mean
  of Equation (\ref{eq-probvlm}) in terms of $M_{BH}$: $\log
  \hat{M}_{BL} = \beta_l \log L^{BL}_{\lambda} + 2 \log v - 2
  \beta_0$, or equivalently $\hat{M}_{BL} \propto
  L^{\beta_l}_{\lambda,BL} v^2$. The intrinsic uncertainty on the
  broad line mass estimates is set by a combination of the intrinsic
  dispersion in $R$ and at a given $L$, and the uncertainty in using
  the single-epoch line width as an estimate of the broad line gas
  velocity dispersion: $\sigma_{\hat{M}_{BL}} =
  2\sigma_{BL}$. Equation (\ref{eq-probvlm}) describes the statistical
  uncertainty in the broad line mass estimates, and does not account
  for any additional systematic errors
  \citep[e.g.,][]{krolik01,collin06}.

  It is typically the case that one employs multiple emission lines to
  estimate $M_{BH}$, producing black hole mass estimates across a
  broad range of redshifts and luminosities. In our work, we use the
  H$\beta$, Mg II, and C IV emission lines. In order to facilitate the
  use of different emission lines in the BHMF estimation, we introduce
  an indicator variable denoted by $\delta$. Here, $\delta_{\rm
  H\beta} = 1$ if the H$\beta$ line width is available, and
  $\delta_{\rm H\beta} = 0$ if the H$\beta$ line widths is not
  available; $\delta_{MgII}$ and $\delta_{CIV}$ are defined in an
  equivalent manner. For example, if one is using optical spectra,
  then at $z = 0.4$ only the H$\beta$ emission line is available, and
  therefore $\delta_{\rm H\beta} = 1, \delta_{MgII} = 0,$ and
  $\delta_{CIV} = 0$.

  Assuming that the line width distributions for each line are
  independent at a given luminosity and black hole mass, then the
  observed distribution of line widths is the product of Equation
  (\ref{eq-probvlm}) for each individual emission line:
  \begin{eqnarray}
    \lefteqn{p(\log {\bf v}|L,M_{BH},z,\beta) =} \nonumber \\
    & & \left[ N(\log v_{{\rm H}\beta}| \bar{v}_{{\rm H}\beta}, 
      \sigma_{{{\rm H}\beta}}^2) \right]^{\delta_{\rm H\beta}}
    \left[ N(\log v_{\rm MgII} | \bar{v}_{\rm MgII}, 
      \sigma_{\rm MgII}^2) \right]^{\delta_{\rm MgII}}
    \left[ N(\log v_{\rm CIV}| \bar{v}_{\rm CIV}, 
      \sigma_{\rm CIV}^2) \right]^{\delta_{\rm CIV}}
      \label{eq-vlmfull}
  \end{eqnarray}
  Here, the average line width for H$\beta$ is $\bar{v}_{\rm H\beta} =
  \beta_0^{\rm H\beta} - (1 / 2)\beta_l^{\rm H\beta} \log L^{\rm
  H\beta}_{\lambda} + (1/2) \log M_{BH}$, and likewise for Mg II and C
  IV. Here, $L^{\rm H\beta}_{\lambda}$ denotes the value of
  $L_{\lambda}$ that is used to calibrate the broad line mass
  estimates for H$\beta$, typically $L_{\lambda}(5100$\AA$)$.

  \citet{vest06} give equations for calculating broad line mass
  estimates from H$\beta$ and C IV, derived from the most recent
  reverberation mapping data \citep{peter04,kaspi05}, and \citet[][in
  progress]{vest07} give an equation for calculating a broad line mass
  estimate from Mg II. These mass scaling relationships are:
  \begin{eqnarray}
    \log \hat{M}_{\rm H\beta} & = & -21.09 + 0.50 \log \lambda L_{\lambda}(5100\AA) + 
    2 \log FWHM_{\rm H\beta} \label{eq-hbblest} \\
    \log \hat{M}_{\rm MgII} & = & -21.21 + 0.50 \log \lambda L_{\lambda}(2100\AA) + 
    2 \log FWHM_{\rm MgII} \label{eq-mgblest} \\
    \log \hat{M}_{\rm CIV} & = & -22.66 + 0.53 \log \lambda L_{\lambda}(1350\AA) + 
    2 \log FWHM_{\rm C IV} \label{eq-c4blest}
  \end{eqnarray}
  For the equations listed above we have used the $FWHM$ of the
  emission line as an estimate of the velocity dispersion, i.e., $v =
  FWHM$. Because $\log \hat{M}_{BL} = \beta_l \log \lambda
  L_{\lambda}^{BL} + 2 \log v - 2 \beta_0$, it follows that
  $\beta^{\rm H\beta}_0 = 10.55, \beta^{\rm MgII}_0 = 10.61, \beta^{C
    IV}_0 = 11.33$, and $\beta_l \approx 0.5$ for all three emission
  lines. In addition, \citet{vest06} find the statistical uncertainty
  in the broad line mass estimates to be 0.43 dex and 0.36 dex for
  H$\beta$ and C IV, respectively. Therefore, since $\sigma_{BL} =
  \sigma_{\hat{M}_{BL}} / 2$, if follows that $\sigma_{\rm H\beta}
  \approx 0.22$ and $\sigma_{\rm C IV} \approx 0.18$ dex. Likewise,
  \citet[][in progress]{vest07} find the intrinsic uncertainty in the
  broad line mass estimate for Mg II to be $\sim 0.4$ dex, and
  therefore $\sigma_{\rm MgII} \approx 0.2$ dex. However, this
  statistical uncertainty may be smaller if a correction is made in
  the virial relationship for radiation pressure \citep{marconi08}.

  Broad line mass estimates are now fairly well understood, and we
  derive our prior distribution for $\beta$ from the scaling results
  of \citet{vest06} and \citet[][in progress]{vest07}. We fix $\beta_l
  = 0.5, 0.5,$ and $0.53$ for H$\beta$, Mg II, and C IV,
  respectively. However, in order to account for the uncertainty in
  these scaling relationships, we consider $\beta_0$ and $\sigma_{BL}$
  to be free parameters in our model.  We cannot estimate the
  normalization and statistical uncertainty in the broad line mass
  estimates solely from the distribution of ${\bf v}, L,$ and $z$,
  since $\beta_0$ and $\sigma_{BL}$ are degenerate with the other
  parameters. Therefore, it is necessary to place constraints on
  $\beta_0$ and $\sigma_{BL}$ through a prior distribution. This
  allows us to constrain $\beta_0$ and $\sigma_{BL}$ while still
  incorporating their uncertainty. The parameters for the prior
  distribution of $\beta_0$ and $\sigma_{BL}$ are based on the
  uncertainty in the scaling relationships of \citet{vest06} and
  \citet[][in progress]{vest07}. Our prior for $\beta_0$ are
  independent Gaussian distributions with means equal to 10.55, 10.61,
  and 11.33 for H$\beta$, Mg II, and C IV, respectively, and standard
  deviations equal to 0.1. To allow greater flexibility in our model,
  we chose the prior standard deviation of $0.1$ to be wider than the
  formal uncertainty on the scaling factors of $\approx 0.02$ reported
  by \citet{vest06}. For each emission line, our prior for
  $\sigma_{BL}$ is a scaled inverse-$\chi^2$ distribution with $\nu =
  25$ degrees of freedom and scale parameter equal to $0.2$ dex. We
  chose $\nu = 25$ degrees of freedom because approximately 25 AGN
  were used to derive the scaling relationships in \citet{vest06}. The
  values of $\beta_0$ were constrained to be within $\pm 0.3$ (i.e.,
  $\pm 3\sigma$) of the values reported by \citet{vest06} and
  \citet[][in progress]{vest07}, and the values of $\sigma_{BL}$ were
  constrained to be within the inverval containing $99\%$ of the
  probability for the scaled inverse-$\chi^2$ distribution. By placing
  these constraints on $\beta_0$ and $\sigma_{BL}$, we ensure that
  their values remain consistent with the results derived from
  reverberation mapping.

  \subsection{Likelihood function for Mixture of Gaussian Functions Model}

  \label{s-mixlik}

  Now that we have formulated the conditional distributions, we can
  calculate the likelihood function for the mixture of Gaussian functions model
  of $\phi(M_{BH},z)$. Comparison with Equation (\ref{eq-thetapost})
  suggests that we need two terms: $p({\bf v}_i,L_{\lambda,i},z_i|\theta)$ and
  $p(I=1|\theta)$. The first term is the joint distribution of line
  widths, luminosities, and redshifts:
  \begin{equation}
    p({\bf v}_i,L_{\lambda,i},z_i|\theta) = \int p({\bf v}_i|L_{\lambda,i},M_{BH,i},\beta)
    p(L_{\lambda,i}|M_{BH,i},\alpha) p(M_{BH,i},z_i|\pi,\mu,\Sigma)\ dM_{BH,i},
    \label{eq-probvlz}
  \end{equation}
  where $\theta = (\alpha, \beta, \pi, \mu, \Sigma)$.

  The integral in Equation (\ref{eq-probvlz}) can be done analytically
  by inserting Equations (\ref{eq-mixmod}), (\ref{eq-problm}), and
  (\ref{eq-vlmfull}) into Equation (\ref{eq-probvlz}). However, the
  result depends on the number of emission lines used for the $i^{th}$
  source. Expressing the likelihood function for a single emission
  line in terms of logarithms, $p(\log v_i,\log L_{\lambda,i},\log
  z_i|\theta)$ is a mixture of $K$ 3-dimensional Gaussian functions:
  \begin{eqnarray}
    p(\log v_i,\log L_{\lambda,i},\log z_i|\theta) & = & \sum_{k=1}^K \frac{\pi_k}{\sqrt{8\pi^3 |V_{k}|}}
    \exp\left\{-\frac{1}{2}({\bf x}_i - \xi_{k})^T V_{k}^{-1} ({\bf x}_i - \xi_{k})\right\} 
      \label{eq-mixlik1} \\
    {\bf x}_i & = & (\log v_i,\log L_{\lambda,i},\log z_i) \label{eq-x1} \\
    \xi_k & = & (\bar{v}_k, \bar{l}_k, \mu_{z,k}) \label{eq-xi1} \\
    \bar{l}_k & = & \alpha_0 + \alpha_m \mu_{m,k} \label{eq-lbar} \\
    \bar{v}_k & = & \beta_0 - \frac{1}{2} \beta_l \bar{l}_{BL,k} + \frac{1}{2} \mu_{m,k} 
    \label{eq-vbar} \\
    \bar{l}_{BL,k} & = & \bar{l}_k + (1 + \alpha_{\lambda}) \log 
    \left(\frac{\lambda_{BL}}{\lambda_{ML}}\right) 
      \label{eq-lbarbl} \\
    V_k & = & \left( \begin{array}{ccc}
      Var(\log v|k) & Cov(\log v,\log l|k) & Cov(\log v,\log z|k) \\
      Cov(\log v,\log l|k) & Var(\log l|k) & \alpha_m \sigma_{mz,k} \\
      Cov(\log v,\log z|k) & \alpha_m \sigma_{mz,k} & \sigma^2_{z,k}
    \end{array} \right) \label{eq-vmat1} \\
    Var(\log v|k) & = & \sigma^2_{BL} + \frac{1}{4}[\beta^2_l Var(\log l|k) + 
      (1 - \alpha_m) \sigma^2_{m,k}] \label{eq-vvar} \\
    Var(\log l|k) & = & \sigma^2_l + \alpha_m^2 \sigma_{m,k}^2 \label{eq-lvar} \\
    Cov(\log v,\log l|k) & = & \frac{1}{2} \alpha_m \sigma^2_{m,k} - \frac{1}{2} \beta_l 
          Var(\log l|k) \label{eq-lvcov} \\
    Cov(\log v,\log z|k) & = & \left( \frac{1}{2} - \frac{1}{2}
      \beta_l \alpha_m \right) \sigma_{mz,k} 
    \label{eq-vzcov}.
  \end{eqnarray}
  Here, $\xi_k$ and $V_k$ are the mean vector and covariance matrix of
  $(\log v_i,\log L_{\lambda,i},\log z_i)$ for the $k^{\rm th}$
  Gaussian function, respectively. In addition, $\bar{l}_k$ is the
  mean $\log L_{\lambda}$ for Gaussian function $k$, $\bar{v}_k$ is
  the mean $v$ for Gaussian function $k$, $\bar{l}_{BL}$ is the mean
  $\log L_{\lambda}^{BL}$ for Gaussian function $k$, $Var(\log v|k)$
  is the variance in $\log v$ for Gaussian function $k$, $Var(\log
  L_{\lambda}|k)$ is the variance in $\log L_{\lambda}$ for Gaussian
  function $k$, $Cov(\log v,\log L_{\lambda}|k)$ is the covariance
  between $\log v$ and $\log L_{\lambda}$ for Gaussian function $k$,
  and $Cov(\log v, \log z|k)$ is the covariance between $\log v$ and
  $\log z$ for Gaussian function $k$; note that $\alpha_m \sigma_{mz,k}$ is the
  covariance between $\log L_{\lambda}$ and $z$ for Gaussian function $k$. The
  mean $\log L_{\lambda}^{BL}$ for Gaussian function $k$ is calculated from
  $\bar{l}_k$ assuming a power-law continuum of the form
  $L_{\lambda}^{BL} = L_{\lambda} (\lambda_{BL} /
  \lambda_{ML})^{\alpha_{\lambda}}$, where $\lambda_{BL}$ is the
  wavelength used in the $R$--$L^{BL}_{\lambda}$ relationship for the
  emission line of interest, and $\lambda_{ML}$ is the wavelength that
  the $M_{BH}$--$L_{\lambda}$ is formulated in. For example,
  $\lambda_{BL} = 5100$\AA\ for the H$\beta$-based mass scaling
  relationship of \citet{vest06}, and $\lambda_{ML}$ may be, say,
  equal to $2500$\AA. Note that we are assuming that
  $\alpha_{\lambda}$ is known.

  In Equation (\ref{eq-mixlik1}) it should be understood that $v_i,
  \beta_0, \beta_l,$ and $\sigma^2_{BL}$ correspond to the particular
  emission line being used. For example, if one is using the C IV line
  width for the $i^{\rm th}$ source, then $v_i = v_{C IV,i}, \beta_0 =
  \beta^{CIV}_0, \beta_l = \beta_l^{CIV},$ and $\sigma^2_{BL} =
  \sigma^2_{CIV}$.
  
  If there are two emission line widths available for the $i^{\rm th}$
  AGN, then $p({\bf v}_i,L_{\lambda,i},z_i|\theta)$ is a mixture of $K$
  4-dimensional Gaussian functions:
  \begin{eqnarray}
    p(\log {\bf v}_i,\log L_{\lambda,i},z_i|\theta) & = & \sum_{k=1}^K \frac{\pi_k}{\sqrt{16\pi^4 |V_{k}|}}
    \exp\left\{-\frac{1}{2}({\bf x}_i - \xi_{k})^T V_{k}^{-1} ({\bf x}_i - \xi_{k})\right\} \label{eq-mixlik2} \\
    {\bf x}_i & = & (\log v_{1,i},\log v_{2,i},\log L_{\lambda,i},\log z_i) \\
    \xi_k & = & (\bar{v}_{1,k}, \bar{v}_{2,k}, \bar{l}_k, \mu_{z,k}) \\
    Cov(\log v_1,\log v_2|k) & = & \frac{1}{4} \left(\beta_{l,1} \beta_{l,2} Var(\log L_{\lambda}|k) + 
    \sigma_m^2 \right)
  \end{eqnarray}
  Here, $Cov(\log v_1,\log v_2|k)$ denotes the covariance between the
  logarithms of the two line widths, $v_1$ and $v_2$, for the $k^{\rm
  th}$ Gaussian function. The $4 \times 4$ covariance matrix of $(\log {\bf
  v}_i, \log L_{\lambda,i}, \log z_i)$ is
  \begin{equation}
    V_k = \left( \begin{array}{cccc}
      Var(\log v_1|k)  & Cov(\log v_1,\log v_2|k)  & Cov(\log v_1,\log L_{\lambda}|k)  & Cov(\log v_1,\log z|k) \\
      Cov(\log v_1,\log v_2|k) & Var(\log v_2|k)   & Cov(\log v_2,\log L_{\lambda}|k)  & Cov(\log v_2,\log z|k) \\
      Cov(\log v_1,\log L_{\lambda}|k)   & Cov(\log v_2,\log L_{\lambda}|k)  & 
      Var(\log L_{\lambda}|k)          & \alpha_m \sigma_{mz,k} \\
      Cov(\log v_1,\log z|k)   & Cov(\log v_2,\log z|k) & \alpha_m \sigma_{mz,k} & \sigma^2_{z,k}
    \end{array} \right).
  \end{equation}
  The other terms are given by Equations
  (\ref{eq-vbar})--(\ref{eq-vzcov}), where it should be understood
  that $\beta_0, \beta_l,$ and $\sigma_{BL}^2$ correspond to the
  values appropriate for each emission line. For example, at $z \sim
  0.6$ both H$\beta$ and Mg II are observable in the optical spectral
  region, and thus it is possible to have line widths for both
  emission lines. In this case, $v_{1,i}$ is the logarithm of the
  H$\beta$ width for the $i^{\rm th}$ source, $v_{2,i}$ is the
  logarithm of the Mg II width for the $i^{\rm th}$ source,
  $\beta_{l,1}$ corresponds to $\beta_l$ for the H$\beta$ line, and
  $\beta_{l,2}$ corresponds to $\beta_l$ for the Mg II line. The
  labeling of the H$\beta$ line width as $v_1$ is irrelevant, and the
  same result would be obtained if we had labeled the H$\beta$ line
  width as $v_2$.

  It should be noted that in Equation (\ref{eq-mixlik2}) we have made
  the assumption that if \emph{at least} one emission line has
  $v_{min} < v < v_{max}$, then $v$ is estimated for all emission
  lines in the observable spectral range at that redshift. If this is
  not the case, then Equation (\ref{eq-mixlik2}) must be integrated
  over $v_{1,i}$ or $v_{2,i}$ if either of $v_{1,i}$ or $v_{2,i}$ fall
  outside of $(v_{min},v_{max})$.

  The term $p(I=1|\theta)$ is the probability that a source is
  included in one's sample for a given set of model parameters
  $\theta$. Under the mixture of Gaussian functions model, Equation
  (\ref{eq-selprob}) can be simplified, allowing more efficient
  calculation. However, as above, the actual functional form of
  $p(I=0|\theta)$ depends on the number of emission lines used in
  broad line mass estimation. If only one emission line is used, then
  Equation (\ref{eq-selprob}) becomes
  \begin{eqnarray}
    p(I=1|\theta) & = & \int_{-\infty}^{\infty} \int_{z_{min}}^{z_{max}} \frac{s(L_{\lambda},z)}{z \ln 10}
    \sum_{k=1}^K \pi_k f_{\bf v}(L_{\lambda},z,\theta,k) 
    N_2({\bf y}_{lz}|\bar{\bf y}_{lz,k}, V_{lz,k})\ dz\ dL_{\lambda}
    \label{eq-mixdetprob} \\
    {\bf y}_{lz} & = & (\log L_{\lambda}, \log z) \\
    \bar{\bf y}_{lz,k} & = & (\bar{l}_k, \mu_{z,k}) \label{eq-lzmu} \\
    V_{lz,k} & = & \left( \begin{array}{cc}
      Var(\log L_{\lambda}|k)              & \alpha_m \sigma_{mz,k} \\
      \alpha_m \sigma_{mz,k} & \sigma^2_{z,k}
    \end{array} \right). \label{eq-lzvar}
  \end{eqnarray}
  The term $f_{\bf v}(L_{\lambda},z,\theta,k)$ is the probability that
  a source has at least one line width between $v_{min}$ and $v_{max}$
  for the $k^{\rm th}$ Gaussian function, given its luminosity and
  redshift. For redshifts where only one emission line is used,
  $f_{\bf v}(L_{\lambda},z,\theta,k) = Pr(v_{min} < v <
  v_{max}|L_{\lambda},z,\theta,k)$, where
  \begin{eqnarray}
    Pr(v_{min} < v < v_{max}|L_{\lambda},z,\theta,k) & = & 
    \Phi\left(\frac{\log v_{max} - E(\log v|L_{\lambda},z,k)}{\sqrt{Var(\log v|L_{\lambda},z,k)}}\right) - 
    \Phi\left(\frac{\log v_{min} - E(\log v|L_{\lambda},z,k)}{\sqrt{Var(\log v|L_{\lambda},z,k)}}\right) 
    \label{eq-vdetprob} \\
    E(\log v|l,z,k) & = & \bar{v}_k + {\bf c}_k^T V_{lz,k}^{-1}
      \left({\bf y}_{lz} - \bar{\bf y}_{lz,k} \right) \label{eq-vcmean1} \\
    Var(\log v|L_{\lambda},z,k) & = & Var(\log v|k) - {\bf c}^T_k V_{lz,k}^{-1} {\bf c}^T_k \label{eq-vcvar1} \\
    {\bf c}_k & = & \left[Cov(\log v,\log L_{\lambda}|k), Cov(\log v,\log z|k) \right].
  \end{eqnarray}
  Here, $\Phi(\cdot)$ is the cumulative distribution function of the
  standard normal distribution, $E(\log v|L_{\lambda},z,k)$ is the mean of
  $\log v$ for the $k^{\rm th}$ Gaussian function at a given $L_{\lambda}$ and
  $z$, $Var(\log v|L_{\lambda},z,k)$ is the variance in $\log v$ for
  the $k^{\rm th}$ Gaussian function at a given $L_{\lambda}$ and $z$, and
  ${\bf c}_k$ is a 2-dimensional vector containing the covariances
  between $\log v$ and both $\log L_{\lambda}$ and $\log z$. The
  standard normal cumulative distribution function can be efficiently
  computed using a look-up table, and therefore only two integrals
  need to be calculated numerically in Equation (\ref{eq-mixdetprob}).

  If one is using multiple emission lines for estimating
  $\phi(M_{BH},z)$, then $f_{\bf v}(L_{\lambda},z,\theta,k)$ must be
  modified to account for this. Equation (\ref{eq-vdetprob}) gives the
  probability that an emission line has a line width $v_{min} < v <
  v_{max}$, under the assumption that only one emission line is used
  at any given redshift. However, if there are redshifts where two
  emission lines are used, then $f_{\bf v}(L_{\lambda},z,\theta,k)$ must be
  modified, as in these cases we need the probability that \emph{at
  least} one emission line has $v_{min} < v < v_{max}$. At redshifts
  where two emission lines are used, $f_{\bf v}(L_{\lambda},z,\theta,k)$ becomes
  the probability that either $v_{min} < v_1 < v_{max}$ or $v_{min} <
  v_2 < v_{max}$:
  \begin{eqnarray}
    \lefteqn{ f_{\bf v}(L_{\lambda},z,\theta,k) = 
      Pr(v_{min} < v_1 < v_{max}|L_{\lambda},z,\theta,k) + Pr(v_{min} < v_2 < v_{max}|L_{\lambda},z,\theta,k)} \\
    & & - Pr(v_{min} < v_1 < v_{max}|L_{\lambda},z,\theta,k)Pr(v_{min} < v_2 < v_{max}|L_{\lambda},z,\theta,k),
    \label{eq-vdetprob2}
  \end{eqnarray}
  where $Pr(v_{min} < v_j < v_{max}|L_{\lambda},z,\theta,k)$ are given by
  Equation (\ref{eq-vdetprob}) for $j=1,2$, respectively.

  As an example, at $z \sim 0.2$ only the H$\beta$ line is available
  in the optical spectral region, and thus, at this redshift, an
  optical survey can only employ the H$\beta$ line for estimating the
  BHMF. In this case, $p(\log v_i,\log L_{\lambda,i},\log z_i|\theta)$ is given
  by Equation (\ref{eq-mixlik1}), and $f_{\bf v}(L_{\lambda},z,\theta,k)$ is
  given by Equation (\ref{eq-vdetprob}). However, at $z \sim 0.6$,
  both H$\beta$ and Mg II are observable in the optical spectral
  region, and thus both may be employed for estimating the BHMF. At
  this redshift, $p(\log {\bf v}_i,\log L_{\lambda,i},\log z_i|\theta)$ is given
  by Equation (\ref{eq-mixlik2}), and $f_{\bf v}(L_{\lambda},z,\theta,k)$ is
  given by Equation (\ref{eq-vdetprob2}), where ${\bf v} = (v_1,v_2)$,
  $v_1$ is the H$\beta$ line width, and $v_2$ is the Mg II line width
  (or vice versa). If only one emission line is available at any
  particular redshift, either because of limited spectral range or
  because of a choice on the part of the researcher to ignore certain
  emission lines, then only Equations (\ref{eq-mixlik1}) and
  (\ref{eq-vdetprob}) need be used.

  The functional forms of $p({\bf v}_i,L_{\lambda,i},z_i|\theta)$ and
  $p(I=1|\theta)$ given above can be inserted into Equation
  (\ref{eq-blobslik}) to obtain the likelihood function for the
  mixture of normals model. A maximum-likelihood estimate of
  $\phi(M_{BH},z)$ can be obtained by first maximizing Equation
  (\ref{eq-blobslik}) with respect to $N$ and $\theta = (\alpha_0,
  \alpha_m, \sigma_l^2, \beta_0, \beta_l, \sigma^2_{BL}, \pi, \mu,
  \Sigma)$. Then, using the maximum-likelihood estimate of $(N, \pi,
  \mu, \Sigma)$, the maximum-likelihood estimate of $\phi(M_{BH},z)$
  is calculated by using Equation (\ref{eq-mixmod}) in Equation
  (\ref{eq-phiconvert}). Unfortunately, for $K>1$ Gaussian functions,
  maximizing the likelihood for the Gaussian mixture model is a
  notoriously difficult optimization problem. The maximization is
  probably most efficiently performed using the
  Expectation-Maximization \citep[EM,][]{em} algorithm, or employing a
  stochastic search routine. Since we focus on Bayesian inference, a
  derivation of the EM algorithm for the BHMF is beyond the scope of
  this work.

  The posterior distribution of $\theta$ and $N$ can be calculated
  using the forms given above for $p(\log {\bf v}_i,\log
  L_{\lambda,i},z_i|\theta)$ and $p(I=1|\theta)$. In this case, one inserts the
  equations for $p(\log {\bf v}_i,\log L_{\lambda,i},\log z_i|\theta)$ and
  $p(I=1|\theta)$ for the Gaussian mixture model into Equations
  (\ref{eq-thetapost}) and (\ref{eq-npost}). The prior distribution,
  $p(\theta)$, is given by Equation (21) in KFV08.

  \subsection{Accounting for Measurement Error}

  \label{s-measerr}

  The preceding discussion has assumed that ${\bf v}_i$ and
  $L_{\lambda,i}$ are known. However, in general, both quantities are
  measured with error. The effect of measurement error is to
  artificially broaden the observed distributions of ${\bf v}_i$ and
  $L_{\lambda,i}$. Because the Bayesian approach attempts to define
  the set of BHMFs that are consistent with the observed distribution
  of ${\bf v}_i, L_{\lambda,i}$, and $z_i$, where `consistency' is
  measured by the posterior probability distribution, measurement
  error can affect statistical inference on the BHMF. If the variance
  of the measurement errors on ${\bf v}_i$ and $L_{\lambda,i}$ are
  small compared to the intrinsic physical variance in these
  quantities, then measurement error does not have a significant
  effect on the results. In general, the measurement errors on
  $L_{\lambda,i}$ will likely be small compared to the physical range
  in AGN luminosities, so we neglect them. This may not always be the
  case for the emission line widths, and in this section we modify the
  likelihood function for the mixture of Gaussian functions model to
  include measurement errors in ${\bf v}_i$. The general method of
  handling measurement errors within a Bayesian or likelihood function
  approach is described in many references
  \citep[e.g.,][]{kelly07a}. For the sake of brevity, we omit the
  derivations and simply report the modifications to the likelihood
  function.

  If one is only employing one emission line at a given redshift, then
  Equation (\ref{eq-probvlz}) can be factored as
  \begin{eqnarray}
    p(\log v_i, \log L_{\lambda,i}, \log z_i|\theta) & = & p(\log v_i|L_{\lambda,i},z_i,\theta)
    p(\log L_{\lambda,i},\log z_i|\theta) \nonumber \\  
    & = & \sum_{k=1}^K \pi_k p(\log v_i|L_{\lambda,i},z_i,\theta,k) 
    p(\log L_{\lambda,i},\log z_i|\theta,k) \label{eq-probvlz2}
  \end{eqnarray}
  Under the mixture of Gaussian functions model, the joint
  distribution of luminosity and redshift for the $k^{th}$ Gaussian
  function is obtained from Equations
  (\ref{eq-mixlik1})--(\ref{eq-vmat1}) by simply omitting the terms
  that depend on $v_i$:
  \begin{equation}
    p(\log L_{\lambda,i},\log z_i|\theta,k) = \frac{1}{\sqrt{4\pi^2 |V_{lz,k}|}}
    \exp\left\{-\frac{1}{2}({\bf y}_{lz,i} - \bar{\bf y}_{lz,k})^T V_{lz,k}^{-1} 
    ({\bf y}_{lz,i} - \bar{\bf y}_{lz,k})\right\}.
    \label{eq-mixliklz}
  \end{equation}
  Here, ${\bf y}_{lz,i} = (\log L_{\lambda,i}, \log z_i)$, $\bar{\bf
    y}_{lz,k}$ is given by Equation (\ref{eq-lzmu}) and $V_{lz,k}$ is
  given by Equation (\ref{eq-lzvar}). The distribution of the measured
  $\log v_i$ at $L_{\lambda,i}$ and $z_i$ for the $k^{th}$ Gaussian
  function is
  \begin{equation}
    p(\log v_i|L_{\lambda,i},z_i,\theta,k) = \frac{1}{\sqrt{2\pi [Var(\log v|L_{\lambda,i},z_i,k) + \sigma^2_{v,i}]}}
    \exp\left\{-\frac{1}{2}\frac{(\log v_i - E(\log v|L_{\lambda,i},z_i,k))^2}{Var(\log v|L_{\lambda,i},z_i,k) + \sigma^2_{v,i}}\right\}.
    \label{eq-cprobv}
  \end{equation}
  Here, $\sigma^2_{v,i}$ is the variance of the measurement error on
  $v_i$, $E(\log v|L_{\lambda,i},z_i,k)$ is given by Equation
  (\ref{eq-vcmean1}), and $Var(\log v|L_{\lambda,i},z_i,k)$ is given
  by Equation (\ref{eq-vcvar1}). From Equation (\ref{eq-cprobv}) the
  effect of measurement error on the line width becomes apparent: the
  distribution of line widths at a given luminosity and redshift is
  broadened by an amount dependent on the magnitude of the line width
  measurement error. If $\sigma^2_{v,i} \ll Var(\log
  v|L_{\lambda,i},z_i,k)$ then Equation (\ref{eq-probvlz2}) reduces to
  Equation (\ref{eq-probvlz}). Otherwise, if measurement error on
  $v_i$ is a concern, Equations (\ref{eq-probvlz2})--(\ref{eq-cprobv})
  should be used for Equation (\ref{eq-probvlz}) instead of Equation
  (\ref{eq-mixlik1}).

  If one is employing two emission lines at a given redshift, then
  Equation (\ref{eq-probvlz}) becomes
  \begin{equation}
    p(\log {\bf v}_i, \log L_{\lambda,i}, \log z_i|\theta) = p(\log v_{1,i}|L_{\lambda,i},z_i,\theta) 
    p(\log v_{2,i}|L_{\lambda,i},z_i,\theta) p(\log L_{\lambda,i},\log z_i|\theta).  \label{eq-probvlz3}
  \end{equation}
  In this case, $p(\log v_{j,i}|L_{\lambda,i},z_i,\theta), j = 1,2,$
  must be calculated seperately for each emission line from Equation
  (\ref{eq-cprobv}).

  \section{POSTERIOR DISTRIBUTION OF THE BHMF VIA MARKOV CHAIN MONTE CARLO}

  \label{s-mha}
  
  The number of free parameters in our statistical model is $6K + 8$,
  where $K$ is the number of Gaussian functions used to approximate $\phi(\log
  M_{BH},\log z)$. Because of the large number of free parameters,
  summarizing the posterior is most efficiently done by using Markov
  Chain Monte Carlo techniques to simulate random draws of $\theta$
  and $N$ from the posterior distribution. In this work we use the
  Metropolis-Hastings algorithm \citep[MHA,][]{metro49,metro53,hast70}
  to perform the MCMC. We use the MHA to obtain a set of random draws
  from the marginal posterior distribution of $\theta$, given by
  Equation (\ref{eq-thetapost}). Then, given the values of $\theta$,
  random draws for $N$ may be obtained from the negative binomial
  distribution. A further description of the Metropolis-Hastings
  algorithm is given by KFV08, and our MHA is an extension of the MHA
  described in KFV08. For further details on the MHA see
  \citet{chib95} or \citet{gelman04}.

  As in KFV08, we denote the current value of a parameter by placing a
  $\tilde{}$ over its symbol, and we denote the proposal value by
  placing a $^*$ in the superscript. For example, if one were updating
  $\alpha_0$, then $\tilde{\alpha}_0$ denotes the current value of
  $\alpha_0$ in the random walk, $\alpha_0^*$ denotes the proposed
  value of $\alpha_0$, $\tilde{\theta}$ denotes the current value of
  $\theta$, and $\theta^*$ denotes the proposed value of $\theta$,
  i.e., $\theta^* = (\alpha_0^*, \tilde{\alpha}_m, \tilde{\sigma}^2_l,
  \tilde{\beta}_0, \tilde{\sigma}^2_{BL}, \tilde{\pi}, \tilde{\mu},
  \tilde{\Sigma}, \tilde{\mu}_0, \tilde{A}, \tilde{T})$. Here,
  $\mu_0,A$ and $T$ are the parameters for the prior distribution on
  the mixture of Gaussian functions parameter (see KFV08). In
  addition, for ease of notation we define $x_{obs} = ({\bf v}_{obs},
  L_{obs}, z_{obs})$ to be the set of observable quantities.

  Our adopted Metropolis-Hastings algorithm is as follows:
  \begin{enumerate}
  \item
    \label{i-mstart}
    Start with initial guesses for $\alpha_0, \alpha_m, \sigma^2_l,
    \beta_0, \sigma^2_{BL}, \pi, \mu, \Sigma, \mu_0,$ and $A$.
  \item
    \label{i-alpha}
    Draw a proposal value for $\alpha_0$ and $\alpha_m$ from a
    2-dimensional normal distribution centered at the current values
    of $\alpha_0$ and $\alpha_m$ with set covariance matrix,
    $\Sigma_{\alpha}$. The proposal values of $\alpha_0$ and
    $\alpha_m$ are then simulated as $(\alpha_0^*, \alpha_m^*) \sim
    N_2([\tilde{\alpha}_0, \tilde{\alpha}_m], \Sigma_{\alpha})$. If
    $p(\theta^*|x_{obs}) > p(\tilde{\theta}|x_{obs})$ then set
    $\tilde{\alpha}_0 = \alpha_0^*$ and $\tilde{\alpha}_m =
    \alpha_m^*$. Otherwise, calculate the ratio $r_{\alpha} =
    p(\theta^*|x_{obs}) / p(\tilde{\theta}|x_{obs})$ and draw a random
    number uniformly distributed between 0 and 1, denoted as $u$. If
    $u < r_{\alpha}$ then set $\tilde{\alpha}_0 = \alpha_0^*$ and
    $\tilde{\alpha}_m = \alpha_m^*$. Otherwise, if $u > r_{\alpha}$,
    the values of $\tilde{\alpha}_0$ and $\tilde{\alpha}_m$ remain
    unchanged.
  \item
    \label{i-sigmal}
    Draw a proposal value for $\log \sigma^2_l$ as $\log
    \tilde{\sigma}^2_l \sim N(2 \log \sigma^*_l,
    \sigma^2_{\sigma_l})$, where $\sigma^2_{\sigma_l}$ is some set
    variance. Similar to before, calculate the ratio $r_{\sigma} =
    \sigma^*_l p(\theta^*|x_{obs}) / \tilde{\sigma}_l
    p(\tilde{\theta}|x_{obs})$. Here, the term $\sigma^*_l /
    \tilde{\sigma}_l$ arises because the MHA acceptance rule must be
    corrected for the asymmetry in the log-normal jumping distribution
    used for $\sigma^2_l$. If $r_{\sigma} \geq 1$ then set
    $\tilde{\sigma}_l = \sigma^*_l$, otherwise set $\tilde{\sigma}_l =
    \sigma^*_l$ with probability $r_{\sigma}$. This is done by drawing
    a uniformly distributed random variable as in step \ref{i-alpha}.
  \item
    \label{i-beta}
    Draw a proposal value for $\beta_0$ from a normal distribution
    centered at the current value of $\beta_0$ with set variance,
    $\sigma^2_{\beta}$. If $p(\theta^*|x_{obs}) >
    p(\tilde{\theta}|x_{obs})$ then set $\tilde{\beta}_0 =
    \beta_0^*$. Otherwise, calculate the ratio $r_{\beta} =
    p(\theta^*|x_{obs}) / p(\tilde{\theta}|x_{obs})$ and draw a random
    number uniformly distributed between 0 and 1, denoted as $u$. If
    $u < r_{\beta}$ then set $\tilde{\beta}_0 = \beta_0^*$. Otherwise,
    if $u > r_{\beta}$, then the value of $\tilde{\beta}_0$ remain
    unchanged. If one is employing multiple emission lines to estimate
    the BHMF, then we have found it faster to simulate proposed values
    of $\beta_0$ for each emission line simultaneously from a
    multivariate normal distribution.
  \item
    \label{i-sigmabl}
    Draw a proposal value for $\log \sigma^2_{BL}$ as $\log
    \tilde{\sigma}^2_{BL} \sim N(2 \log \sigma^*_{BL},
    \sigma^2_{\sigma_{BL}})$, where $\sigma^2_{\sigma_{BL}}$ is some
    set variance. Similar to the update for $\sigma^2_l$, calculate
    the ratio $r_{BL} = \sigma^*_{BL} p(\theta^*|x_{obs}) /
    \tilde{\sigma}_{BL} p(\tilde{\theta}|x_{obs})$. If $r_{BL} \geq 1$
    then set $\tilde{\sigma}_{BL} = \sigma^*_{BL}$, otherwise set
    $\tilde{\sigma}_{BL} = \sigma^*_{BL}$ with probability
    $r_{\sigma}$. This is done by drawing a uniformly distributed
    random variable as in step \ref{i-alpha}. If one is employing
    multiple emission lines to estimate the BHMF, then we have found
    it faster to simulate proposed values of $\log \sigma^2_{BL}$ for
    each emission line simultaneously from a multivariate normal
    distribution.
  \item
    \label{i-repeat}
    Draw new values of the Gaussian mixture model parameters according
    to steps 2--6 in the MHA described in KFV08.
  \end{enumerate}
  One then repeats steps \ref{i-alpha}--\ref{i-repeat} until the MCMC
  converges, saving the values of $\tilde{\theta}$ at each
  iteration. After convergence, the MCMC is stopped, and the values of
  $\tilde{\theta}$ may be treated as a random draw from the marginal
  posterior distribution of $\theta$, $p(\theta|x_{obs})$.  Techniques
  for monitering convergence of the Markov Chains can be found in
  \citet{gelman04}. Given the values of $\theta$ obtained from the
  MCMC, one can then draw values of $N$ from the negative binomial
  distribution (cf. Eq.[\ref{eq-npost}]).

  Having obtained random draws of $N$ and $\theta$ from
  $p(\theta,N|{\bf v}_{obs}, L_{obs}, z_{obs})$, one can then use
  these values to calculate an estimate of $\phi(M_{BH},z)$, and its
  corresponding uncertainty. This is done by using each of the
  MCMC draws of $\theta$ and $N$ to calculate Equation
  (\ref{eq-mixbhmf}). The posterior distribution of $\phi(M_{BH},z)$
  can be estimated for any value of $M_{BH}$ and $z$ by plotting a
  histogram of the values of $\phi(M_{BH},z)$ obtained from the MCMC
  values of $\theta$ and $N$. KFV08 illustrates in more detail how to
  use the MHA results to perform statistical inference.

  \section{APPLICATION TO SIMULATED DATA}

  \label{s-sim}

  As an illustration of the effectiveness of our method, we applied it
  to a simulated data set. Because we will eventually apply this
  method to the BHMF for the SDSS DR3 quasar catalogue
  \citep{dr3qsos}, we assume the effective survey area and selection
  function reported for the DR3 quasar sample \citep{dr3lumfunc}.

  \subsection{Construction of the Simulated Sample}

  \label{s-simconst}

  We construct our simulated survey in a manner very similar to that
  used by KFV08. We first drew a random value of $N_{\Omega}$ quasars
  from a binomial distribution with probability of success $\Omega / 4
  \pi = 0.0393$ and number of trials $N = 2 \times 10^5$. Here,
  $\Omega = 1622\ {\rm deg}^2$ is the effective sky area for our
  simulated survey, and we chose the total number of quasars to be $N
  = 2 \times 10^5$ in order to produce a value of $n \sim 1000$
  observed sources after including the flux limit. While this produces
  a much smaller sample than the actual sample of $\sim 1.5 \times
  10^4$ quasars from the SDSS DR3 luminosity function work
  \citep{dr3lumfunc}, we chose to work with this smaller sample to
  illustrate the effectiveness of our method on more moderate sample
  sizes. This first step of drawing from a binomial distribution
  simulates a subset of $N_{\Omega}$ sources randomly falling within
  an area $\Omega$ on the sky, where the total number of sources is
  $N$. Note that we have not included any flux limits yet.

  For each of these $N_{\Omega} \sim 8000$ sources, we simulated
  values of $M_{BH}$ and $z$. We first simulated values of $\log z$
  from a distribution of the form
  \begin{equation}
    g(\log z) = \frac{4\Gamma(a + b)}{\Gamma(a) \Gamma(b)} \frac{\exp(a \zeta^*)}
               {\left(1 + \exp(\zeta^*) \right)^{a+b}}, \label{eq-zmarg}
  \end{equation}
  where $\zeta^* = 4 (\log z - 0.4)$. The parameters $a = 2$ and $b =
  3$ were chosen to give an observed redshift distribution similar to
  that seen for SDSS DR3 quasars \citep[e.g.,][]{dr3lumfunc}.

  For each simulated value of $z$, we simulated a value of $M_{BH}$ by
  taking the distribution of $M_{BH}$ at a given redshift to be a
  smoothly-connected double power-law. In this case, the conditional
  distribution of $\log M_{BH}$ at a given $z$ is
  \begin{eqnarray}
    g(\log M_{BH}|z) & \propto & M_{BH}^{\gamma(z) / \ln 10}
    \left[ 1 + \left(\frac{M_{BH}}{M_{BH}^* (z)}\right)^{(\gamma(z) + \delta(z)) / 
	\ln 10} \right]^{-1} \label{eq-mcond} \\
    \gamma(z) & = & 2.5 + 0.5 \log z \\
    \delta(z) & = & 4.75 + 2 \log z \\
    \log M_{BH}^*(z) & = & 7.5 + 3 \log (1 + z),
  \end{eqnarray}
  where $\log M_{BH}^*(z)$ approximately marks the location of the
  peak in $g(\log M_{BH}|z)$, $\gamma(z)$ is the slope of $\log g(\log
  M_{BH}|z)$ for $M_{BH} \lesssim M_{BH}^*(z)$, and $\delta(z)$ is the
  slope of $\log g(\log M_{BH}|z)$ for $M_{BH} \gtrsim
  M_{BH}^*(z)$. For our simulation, both the peak and logarithmic
  slopes of the BHMF evolve.

\begin{figure}
  \begin{center}
    \includegraphics[scale=0.5,angle=90]{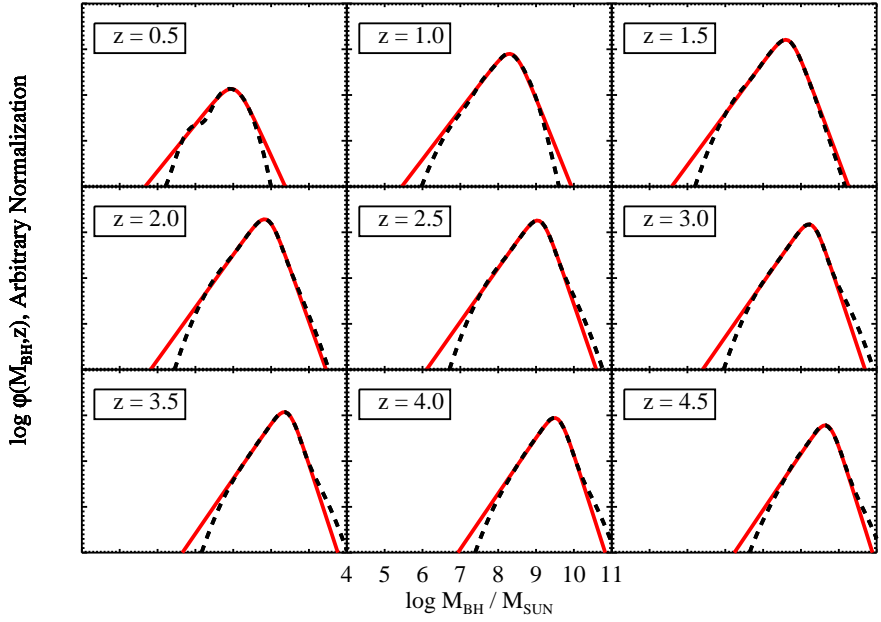}
    \caption{The true BHMF (solid red line) at several values of $z$,
    and the best $K=4$ Gaussian function fit (dashed black line). In
    this case, approximating the BHMF with $K=4$ 2-dimensional
    Gaussian functions provides a good fit. The mixture of Gaussian
    functions approximation diverges from the true BHMF in the tails
    of the distribution of $M_{BH}$. However, in general, the
    uncertainties on the BHMF in the tails are dominated by the
    statistical errors due to the small number of sources in these
    regions, and not by the bias introduced from approximating the
    BHMF as a mixture of Gaussian functions. \label{f-truebhmf}}.
  \end{center}
\end{figure}

  The joint probability distribution of $\log M_{BH}$ and $\log z$ is
  $g(\log M_{BH},\log z) = g(\log M_{BH}|z) g(\log z)$, and therefore
  Equations (\ref{eq-zmarg}) and (\ref{eq-mcond}) imply that the true
  BHMF for our simulated sample is
  \begin{equation}
    \phi_0(M_{BH},z) \propto \frac{N}{z M_{BH}} \left(\frac{dV}{dz}\right)^{-1} 
    g(\log M_{BH}|z) g(\log z).
    \label{eq-bhmftrue}
  \end{equation}
  The constant of proportionality in Equation (\ref{eq-bhmftrue}) can
  be calculated by noting that \\ $\int \int \phi_0(M_{BH},z)\
  dM_{BH}\ dV = N$. Figure \ref{f-truebhmf} shows $\phi_0(M_{BH},z)$
  at several redshifts. Also shown in Figure \ref{f-truebhmf} is the
  best fit for a mixture of $K = 4$ Gaussian functions. Despite the
  fact that $\phi_0(M_{BH},z)$ has a rather complicated parameteric
  form, a mixture of four Gaussian functions is sufficient to achieve
  a good approximation to $\phi_0(M_{BH},z)$.

  For each simulated black hole mass and redshift, we simulated a
  luminosity according to Equation (\ref{eq-mlrel}). However, unlike
  the Gaussian distribution assumed in this work (see
  Eq.[\ref{eq-problm}]), we assume an asymetric distribution of
  Eddington ratios that evolves as $\bar{\Gamma}_{Edd} \propto
  \sqrt{1+z}$. We do this in order to test the robustness of our
  simple assumption that the distribution of $L_{\lambda}$ at a given
  $M_{BH}$ is independent of redshift and given by a normal
  distribution. In this simulated `universe', the distribution of
  $\Gamma_{Edd}$ does not evolve strongly, as is implied by
  observations \citep[e.g.,][]{vest04,koll06}.

\begin{figure}
  \begin{center}
    \includegraphics[scale=0.33,angle=90]{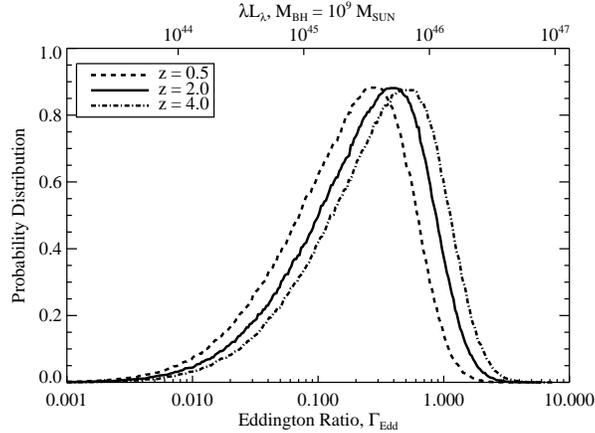}
    \caption{Distribution of Eddington ratios, $\Gamma_{Edd}$, for our
      simulated survey at $z = 0.5, 2,$ and $4$. The corresponding
      values of $\lambda L_{\lambda} [2500$\AA$]$ are shown along the
      top of the plot for a black hole with $M_{BH} = 10^9 M_{\odot}$
      and a bolometric correction of $C_{\lambda} = 10$.
    \label{f-eddrat}}
  \end{center}
\end{figure}

  To simulate values of luminosity at a given black hole mass, we
  first simulated values of the Eddington ratio from a skew-normal
  distribution as
  \begin{equation}
    \log \Gamma_{Edd} = 0.2 \epsilon - 0.75 |\delta| - 0.3 + 0.5 \log (1 + z).
    \label{eq-skewnorm}
  \end{equation}
  Here, $\epsilon$ and $\delta$ are both random deviates independently
  drawn from the standard normal distribution, i.e., $\epsilon, \delta
  \sim N(0,1)$. Figure \ref{f-eddrat} shows the distribution of
  $\Gamma_{Edd}$ at a few different redshifts. Values of $\lambda
  L_{\lambda}$ were then calculated according to Equation
  (\ref{eq-mlrel}) assuming a constant bolometric correction of
  $C_{\lambda} = 10$ \citep[e.g.,][]{kaspi00}. For simplicity, we only
  use a constant bolometric correction for all simulated quasars. In
  our simulation we take $\lambda = 2500$\AA; the choice of $\lambda$
  is arbitrary and has no material effect on our results. The median
  Eddington ratio for our simulated sample is $\Gamma_{Edd} \approx
  0.25$, and the dispersion in $\log \Gamma_{Edd}$ is $\approx 0.5$
  dex. Because the mean $\Gamma_{Edd}$ evolves in our simulation, and
  because the mean $M_{BH}$ evolves, $\Gamma_{Edd}$ and $M_{BH}$ are
  slightly correlated due to the shared correlation with $z$:
  $\Gamma_{Edd} \propto M_{BH}^{0.09}$. Therefore, $L_{\lambda}
  \propto M_{BH}^{1.09}$. Comparison with Equation (\ref{eq-problm})
  suggest that we would expect $\alpha_0 \sim 36, \alpha_m \sim
  1.09,$ and $\sigma_l \sim 0.5$ dex.

  For each simulated black hole mass and luminosity, we simulated
  broad emission line widths for H$\beta$, Mg II, and C IV according
  to Equation (\ref{eq-probvlm}). We simulated values of the H$\beta$
  line width for $0 < z < 0.9$, values of the Mg II line width for
  $0.4 < z < 2.2$, and values of the C IV line width for $1.6 < z <
  4.5$. Note that for this simulation both H$\beta$ and Mg II are
  available at $0.4 < z < 0.9$, and both Mg II and C IV are available
  at $1.6 < z < 2.2$. Based on the most recent reverberation mapping
  data \citep{kaspi05,bentz06}, we took $R \propto L_{\lambda}^{1/2}$ ($\beta_l
  = 0.5$) for all emission lines. In addition, we set $\beta_0 = 10.6,
  10.6,$ and $10.7$ for the H$\beta$, Mg II, and C IV emission lines,
  respectively; these values were chosen to give emission line $FWHM$
  with typical values of several thousand ${\rm km}\ s^{-1}$. The
  dispersion in the logarithm of the emission line width at a given
  luminosity and black hole mass was taken to be $\sigma_{BL} = 0.25,
  0.225,$ and $0.2$ for H$\beta$, Mg II, and C IV, respectively. These
  values of $\sigma_{BL}$ were chosen to give broad line mass estimate
  statistical uncertainties similar to that found from the
  reverberation mapping data \citep{vest06}.
  
  We randomly kept each source, where the probability of including a
  source given its luminosity and redshift was taken to be the SDSS
  DR3 Quasar selection function, as reported by \citet{dr3lumfunc}. In
  addition, we only kept sources with at least one emission line
  having a line width $1000\ {\rm km\ s^{-1}} < v < 1.8 \times 10^4\
  {\rm km\ s^{-1}}$. Sources with $v < 1000$ were assumed to be
  indistinguishable from narrow-line AGN, and sources with $v > 1.8
  \times 10^4$ were assumed to be too difficult to distinguish from
  the underlying continuum and iron emission, and are thus too broad
  to be able to obtain a reliable estimate of the line width. After
  simulating the effects of the selection function, we were left with
  a sample of $n \sim 1000$ sources. Therefore, our simulated survey was
  only able to detect $\sim 0.5\%$ of the $N = 2 \times 10^5$ total
  quasars in our simulated `universe'.

\begin{figure}
  \begin{center}
    \scalebox{0.5}{\rotatebox{90}{\plotone{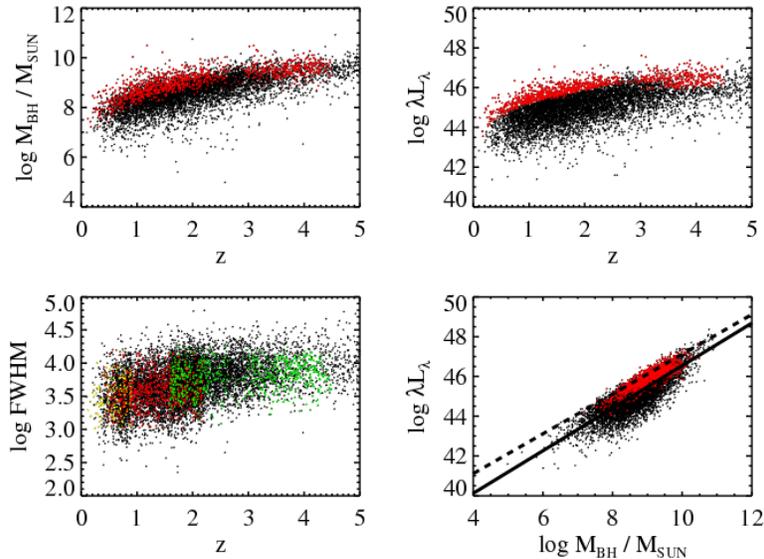}}}
    \caption{The distribution of $M_{BH},L_{\lambda},$ and $FWHM$ for
      our simulated sample. Red dots denote sources included in the
      sample, and black dots denote sources not included in the
      sample. In the plot of $FWHM$ as a function of $z$, yellow dots
      denote sources with $H\beta$ measurements, red dots denote
      sources with Mg II measurements, and green dots denots sources
      with C IV measurements. In the plot of $L_{\lambda}$ as a
      function of $M_{BH}$, the solid line shows the best linear
      regression of $\log L_{\lambda}$ as a function of $\log M_{BH}$,
      and the dashed line shows the Eddington limit for a bolometric
      correction of $C_{\lambda} = 10$. \label{f-simdist}}
  \end{center}
\end{figure}

  The distributions of $M_{BH},z,L_{\lambda},$ and $v$ are shown in Figure
  \ref{f-simdist} for both the detected sources and the full
  sample. As can be seen, the majority of sources are missed by our
  simulated survey, and that the fairly `hard' limit on luminosity
  corresponds to a much `softer' limit on $M_{BH}$. In particular,
  almost all simulated quasars with $M_{BH} \lesssim 10^8 M_{\odot}$
  are missed at $z \gtrsim 1$, and all simulated quasars with $M_{BH}
  \lesssim 10^7 M_{\odot}$ are missed at any redshift.

  To simulate the effects of using values of $\beta_0$ and
  $\sigma_{BL}$ derived from a reverberation mapping sample, we
  simulated a sample of 25 low-$z$ sources with known $M_{BH}$; these
  low-$z$ sources were simulated in the same manner as described
  above. We then used these 25 `reverberation mapping' sources to fit
  $\beta_0$ and $\sigma_{BL}$. The fitted values were then used for
  our prior distribution on $\beta_0$ and $\sigma_{BL}$ as described
  in \S~\ref{s-vprob}.

  \subsection{Performing Statistical Inference on the BHMF with the MCMC Output}

  \label{s-simmcmc}

  We performed the MHA algorithm described in \S~\ref{s-mha} to obtain
  random draws from the posterior distribution for this sample,
  assuming the Gaussian mixture model described in
  \S~\ref{s-smodel}. We performed $10^4$ iterations of burn-in, and
  then ran the markov chains for an additional $3 \times 10^4$. We ran
  five chains at the same time in order to monitor convergence
  \citep[e.g., see][]{gelman04} and explore possible multimodality in
  the posterior. The chains had converged after $4 \times 10^4$ total
  iterations, leaving us with $\sim 1.5 \times 10^5$ random draws
  from the posterior distribution, $p(\theta, N|{\bf v}_{obs},
  L_{obs}, z_{obs})$.

\begin{figure}
  \begin{center}
    \scalebox{0.7}{\rotatebox{90}{\plotone{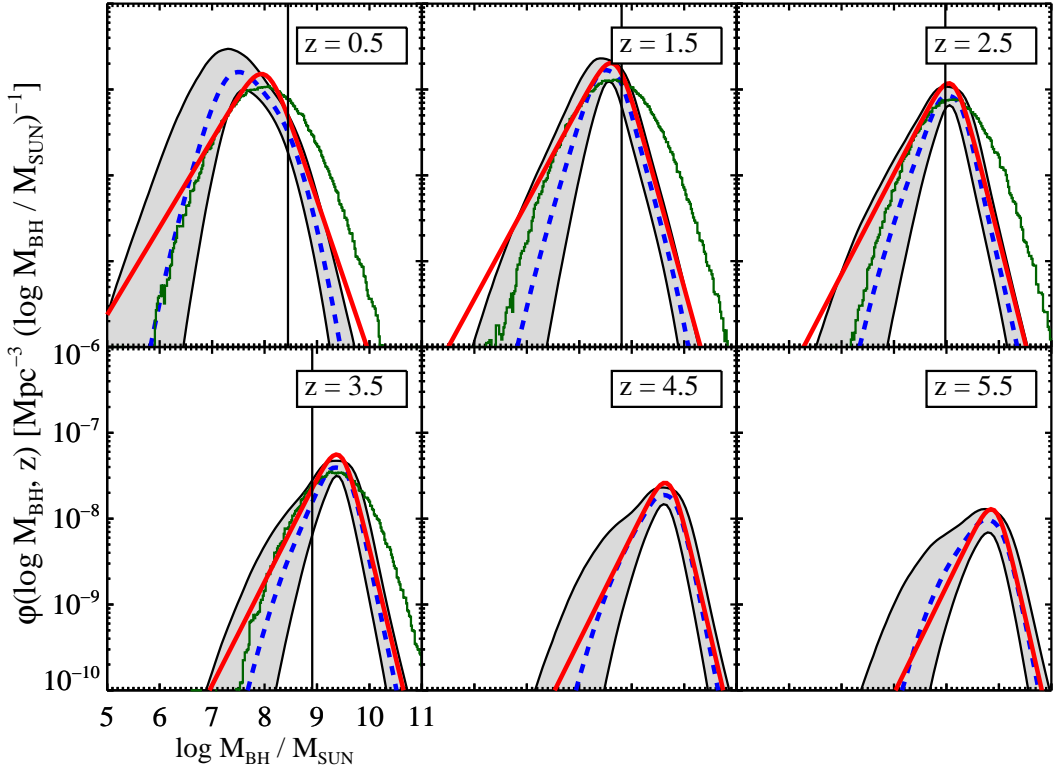}}}
    \caption{The true BHMF (solid red line) at several redshifts. The
      axis labels are the same for all panels, but for clarity we only
      place exterior labels on the bottom left panel. Also shown is
      the posterior median estimate of the BHMF based on the mixture
      of Gaussian functions model (dashed blue line), the region
      containing $68\%$ of the posterior probability (shaded region),
      and the expected value for a $1/V_a$-type binned estimate based
      on the broad emission line estimates, $\hat{\phi}_{BL}$ (thin
      bumpy solid green line). The vertical lines mark the 50\%
      incompleteness limit for a quasar with FWHM = $4000{\rm \ km\
      s^{-1}}$, a typical value for the simulated sources. Note that
      in general the best-fit mixture of Gaussian functions
      approximation to the BHMF will not equal the true BHMF, as it is
      derived from a finite random sample drawn from the true
      BHMF. The bayesian mixture of Gaussian functions model is able
      to accurately constrain the BHMF, even below the survey
      detection limit. However, $\hat{\phi}_{BL}$ provides a biased
      estimate of the BHMF. \label{f-philog}}
  \end{center}
\end{figure}

  In Figure \ref{f-philog} we show $\phi(\log M_{BH},z)$ at several
  different redshifts, on both a linear scale and a logarithmic
  scale. In general, we find it easier to work with $\phi(\log
  M_{BH},z) = \ln 10 M_{BH} \phi(M_{BH},z)$, as $\phi(M_{BH},z)$ can
  span several orders of magnitude in $M_{BH}$. Figure \ref{f-philog}
  shows the true value of the BHMF, $\phi_0(\log M_{BH},z)$, the
  best-fit estimate of $\phi(\log M_{BH},z)$ based on the mixture of
  Gaussian functions model, and the regions containing $68\%$ of the
  posterior probability. Here, as well as throughout this work, we
  will consider the posterior median of any quantity to be the
  `best-fit' for that quantity. In addition, in this work we will
  report errors at the $68\%$ level unless specified otherwise, and
  therefore the regions containing $68\%$ of the posterior probability
  can be loosely interpreted as asymmetric error bars of length
  $\approx 1\sigma$. As can be seen, the true value of $\phi(\log
  M_{BH},z)$ is contained within the $68\%$ probability region for
  most of the values of $\log M_{BH}$, even those below the survey
  detection limit.

  We compare our method with an estimate of the BHMF obtained by
  combining the broad line mass estimates with the more traditional $1
  / V_a$ estimator, developed for luminosity function estimation. We
  do this primarily to illustrate the pitfalls that can arise from
  employing broad line mass estimates and not properly accounting for
  the black hole mass selection function. Following \citet{fan01}, we
  denote the effective volume of the $i^{\rm th}$ source as
  $V_a^i$. If the $i^{\rm th}$ source lies in a redshift bin of width
  $\Delta z$ and has a luminosity $L_{\lambda,i}$, then
  \begin{equation}
    V^i_a = \int_{\Delta z} s(L_{\lambda,i},z) \left(\frac{dV}{dz} \right)\ dz.
    \label{eq-veff}
  \end{equation}
  Dividing up the $(\log M_{BH},z)$ plane into bins of width $\Delta
  \log M_{BH} \times \Delta z$, one may be tempted to calculate an
  estimate of $\phi(\log M_{BH},z)$ based on the broad line estimates
  of $\log M_{BH}$ as
  \begin{equation}
    \hat{\phi}_{BL}(\log M_{BH},z) = \frac{1}{\Delta \log M_{BH}} \sum_i \frac{1}{V_a^i}.
    \label{eq-phihat}
  \end{equation}
  Here, the sum is over all sources with broad lines estimates $\log
  M_{BH} \leq \log \hat{M}_{BL,i} \leq \log M_{BH} + \Delta \log
  M_{BH}$ and $z \leq z_i \leq z + \Delta z$.

  Figure \ref{f-philog} also displays the expected value of
  $\hat{\phi}_{BL}$ for $z = 0.5, 1.5, 2.5, 3.5$ and $4.5$. In order
  to estimate the expected value of $\hat{\phi}_{BL}$ at each $z$, we
  simulated $10^7$ quasars at each redshift interval. This produces
  extremely small error bars on $\hat{\phi}_{BL}$ and allows us to
  estimate the value of $\hat{\phi}_{BL}$ that would be obtained on
  average, i.e., in the limit of an infinitely large sample. As can be
  seen, $\hat{\phi}_{BL}$ is a biased estimate of the BHMF. This bias
  is caused by a combination of the relatively large statistical
  uncertainties on the broad line mass estimates, which broaden the
  inferred BHMF, and by the use of the luminosity selection function
  instead of the black hole mass selection function in the $1 / V_a$
  correction. The large statistical uncertainties on the broad line
  mass estimates broaden the inferred BHMF, and therefore
  $\hat{\phi}_{BL}$ significantly overestimates the BHMF at the high
  mass end, while underestimating the BHMF near its peak. In addition,
  $\hat{\phi}_{BL}$ underestimates the BHMF at the low mass end due to
  the inability of the $1 / V_a$ technique to completely correct for
  incompleteness. The end result is a systematic shift in the inferred
  BHMF toward higher $M_{BH}$, and a similar effect has been noted by
  \citet{shen07}. The effective volume in Equation (\ref{eq-veff}) is
  defined based on the detection probability as a function of
  luminosity, and not black hole mass. As mentioned in
  \S~\ref{s-selfunc}, in order to correctly apply the $1 / V_a$
  estimator for BHMF estimation it is necessary to obtain the black
  hole mass selection function, given by Equation
  (\ref{eq-mselect_convert}). However, this requires knowledge of
  $p(L_{\lambda}|M_{BH},z)$. Furthermore, even if there were no
  selection effects, $\hat{\phi}_{BL}$ would still be biased because
  of the significant uncertainty ($\sim 0.4$ dex) on $\log
  \hat{M}_{BL}$.

\begin{figure}
  \begin{center}
    \scalebox{0.5}{\rotatebox{90}{\plotone{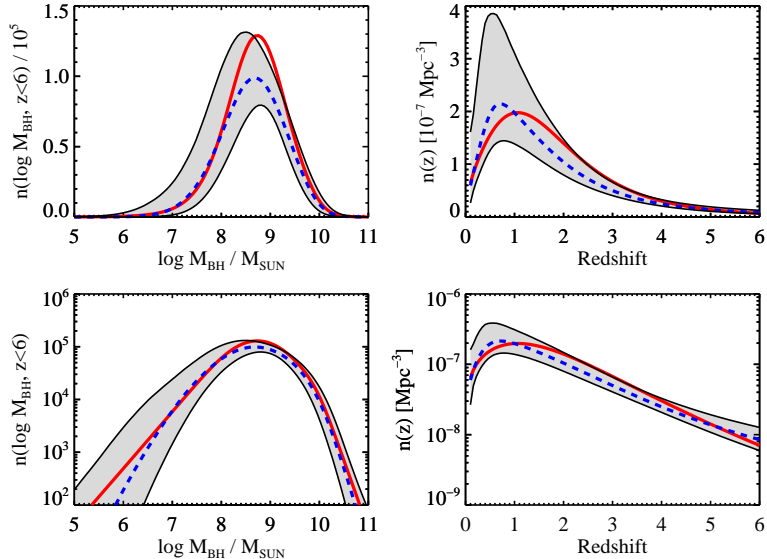}}}
    \caption{The integrated $z < 6$ quasar number density (number per
    $\log M_{BH}$ interval, left two panels) and the comoving quasar
    number density as a function of $z$ (number per ${\rm Mpc}^3$,
    right two panels). The top two panels show a linear stretch and
    the bottom two panels show a logarithmic stretch. As with Figure
    \ref{f-philog}, the solid red line denotes the true value for the
    simulation, the dashed blue line denotes the posterior median for
    the mixture of Gaussian functions model, and the shaded regions
    contain $68\%$ of the posterior probability. The posterior median
    provides a good fit to the true values, and the uncertainties
    derived from the MCMC algorithm based on the Gaussian mixture
    model are able to accurately constrain the true values of these
    quantities, despite the flux limit.\label{f-mzmarg}}
  \end{center}
\end{figure}

  As in KFV08, we can use the MCMC output to constrain various
  quantities of interest calculated from the BHMF.  Figure
  \ref{f-mzmarg} compares the true integrated $z < 6$ number
  distribution of $\log M_{BH}$, $n(\log M_{BH}, z < 6)$, with the
  mixture of Gaussian functions estimate.  The quantity $n(\log
  M_{BH}, z<6) d\log M_{BH}$ is the number of quasars at $z < 6$ with
  black hole masses between $\log M_{BH}$ and $\log M_{BH} + d\log
  M_{BH}$. KFV08 give an equation for calculating $n(\log L,z < z_0)$
  based on the mixture of Gaussian functions model (see their
  Eq.[42]), and $n(\log M_{BH},z < z_0)$ is calculated in an
  equivalent manner. Similar to Figure \ref{f-philog}, the true value
  of $n(\log M_{BH},z < 6)$ is contained within the $68\%$ probability
  region for most values of $M_{BH}$, even those below the survey
  detection limit.

  In addition, in Figure \ref{f-mzmarg} we show the comoving number
  density of broad line AGN as a function of redshift, $n(z)$. This is
  obtained by integrating $\phi(M_{BH},z)$ over all possible values of
  $M_{BH}$, given by Equation (45) of KFV08.  As before, the true
  value of $n(z)$ is contained within the $68\%$ probability region,
  despite the fact that the integration extends over \emph{all}
  $M_{BH}$, even those below the detection limit. The wider confidence
  regions reflect additional uncertainty in $n(z)$ resulting from
  integration over those $M_{BH}$ below the detection limit. In
  particular, the term $dV / dz$ becomes small at low redshift, making
  the estimate of $n(z)$ more unstable as $z \rightarrow 0$, and thus
  inflating the uncertainties at low $z$.

\begin{figure}
  \begin{center}
    \scalebox{0.5}{\rotatebox{90}{\plotone{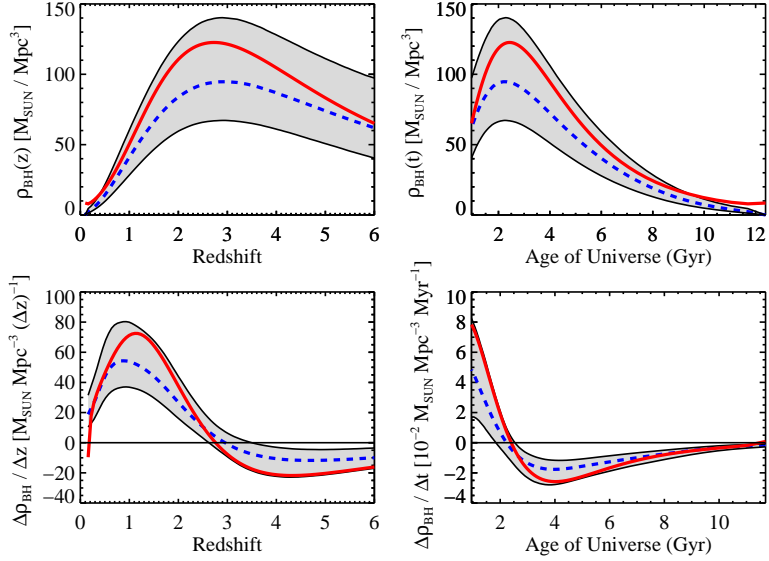}}}
    \caption{Comoving broad line quasar black hole mass density (top
      two panels) and its derivative (bottom two panels), shown as a
      function of redshift (left two panels) and cosmic age (right two
      panels). The plotting symbols are the same as in Figure
      \ref{f-mzmarg}. As in the previous figures, the Gaussian mixture
      model is able to provide an accurate fit to the true values of
      $\rho_{BH}^{QSO}(z)$, and the bayesian MCMC approach is able to
      provide accurate constraints on $\rho_{BH}^{QSO}(z)$ and
      $d\rho_{BH}^{QSO} / dz$, despite the fact that the integral used
      for calculating these quanties extends below the survey
      detection limit. \label{f-rhoz}}
  \end{center} 
\end{figure}

  Two other potentially useful quantities are the comoving black hole
  mass density for quasars, $\rho_{BH}^{QSO}(z)$, and its
  derivative. The comoving black hole mass density is given by
  $\rho_{BH}^{QSO}(z) = \int_0^{\infty} M_{BH} \phi(M_{BH},z)\
  dM_{BH}$. The quantity $\rho_{BH}^{QSO}(z)$ is given by Equation (47)
  of KFV08 and replacing luminosity with black hole mass. We calculate
  the derivative of $\rho_{BH}^{QSO}(z)$ numerically. Figure \ref{f-rhoz}
  compares the true values of $\rho_{BH}^{QSO}(z)$ and its derivative with
  the posterior distribution for $\rho_{BH}^{QSO}(z)$ inferred from the
  mixture model, both as a function of $z$ and the age of the universe
  at redshift $z$, $t(z)$. Comparison with Figure \ref{f-mzmarg}
  reveals that the comoving quasar black hole mass density,
  $\rho_{BH}^{QSO}(z)$, is a better constrained quantity than the comoving
  quasar number density, $n(z)$.  Furthermore, $n(z)$ appears to peak
  later than $\rho_{BH}^{QSO}(z)$. We can correctly infer that the quasar
  comoving black hole mass density reaches it point of fastest growth
  at $t(z) \lesssim 1$ Gyr, and its point of fastest decline at $t(z)
  \sim 4$ Gyr.

\begin{figure}
  \begin{center}
    \includegraphics[scale=0.33,angle=90]{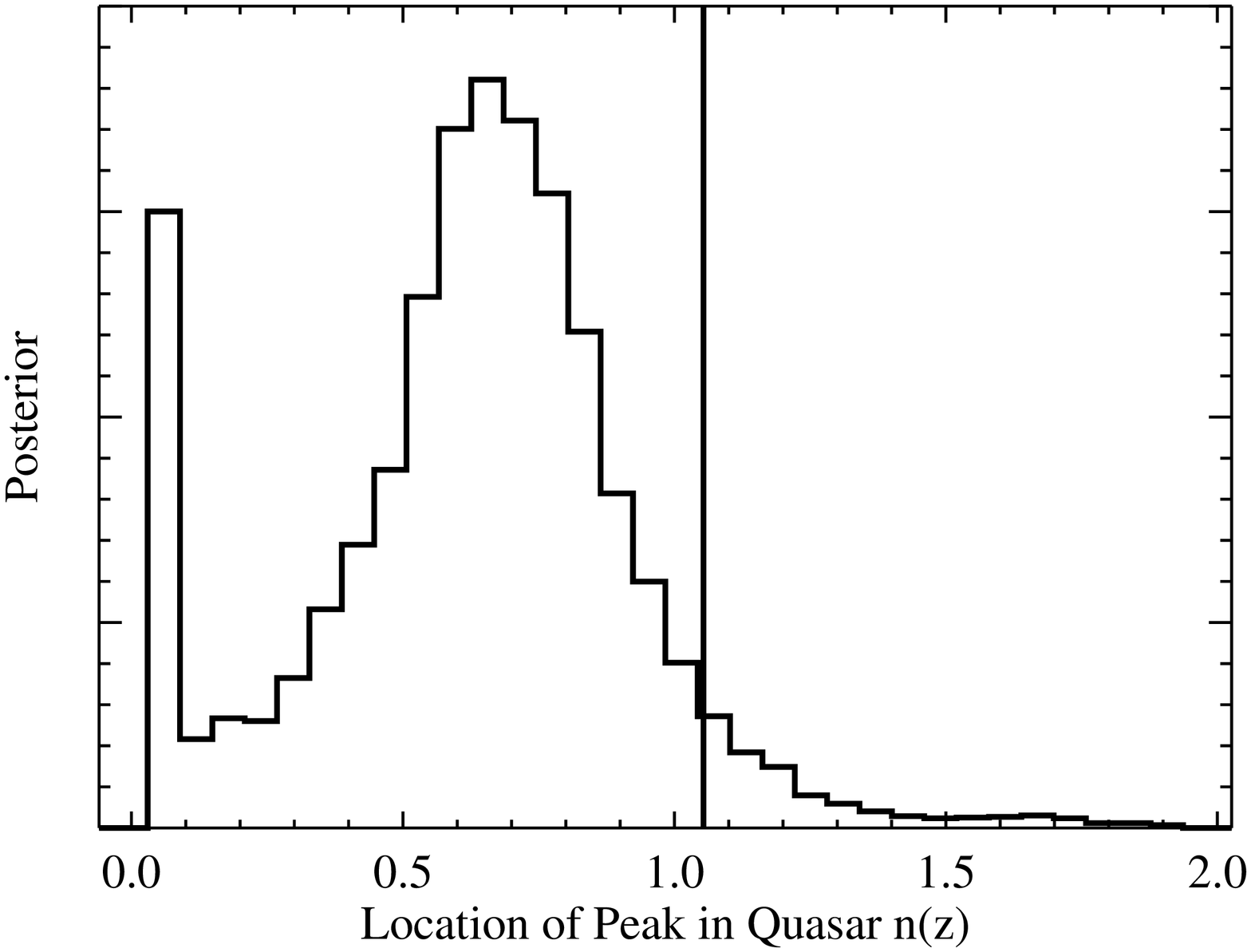}
    \includegraphics[scale=0.33,angle=90]{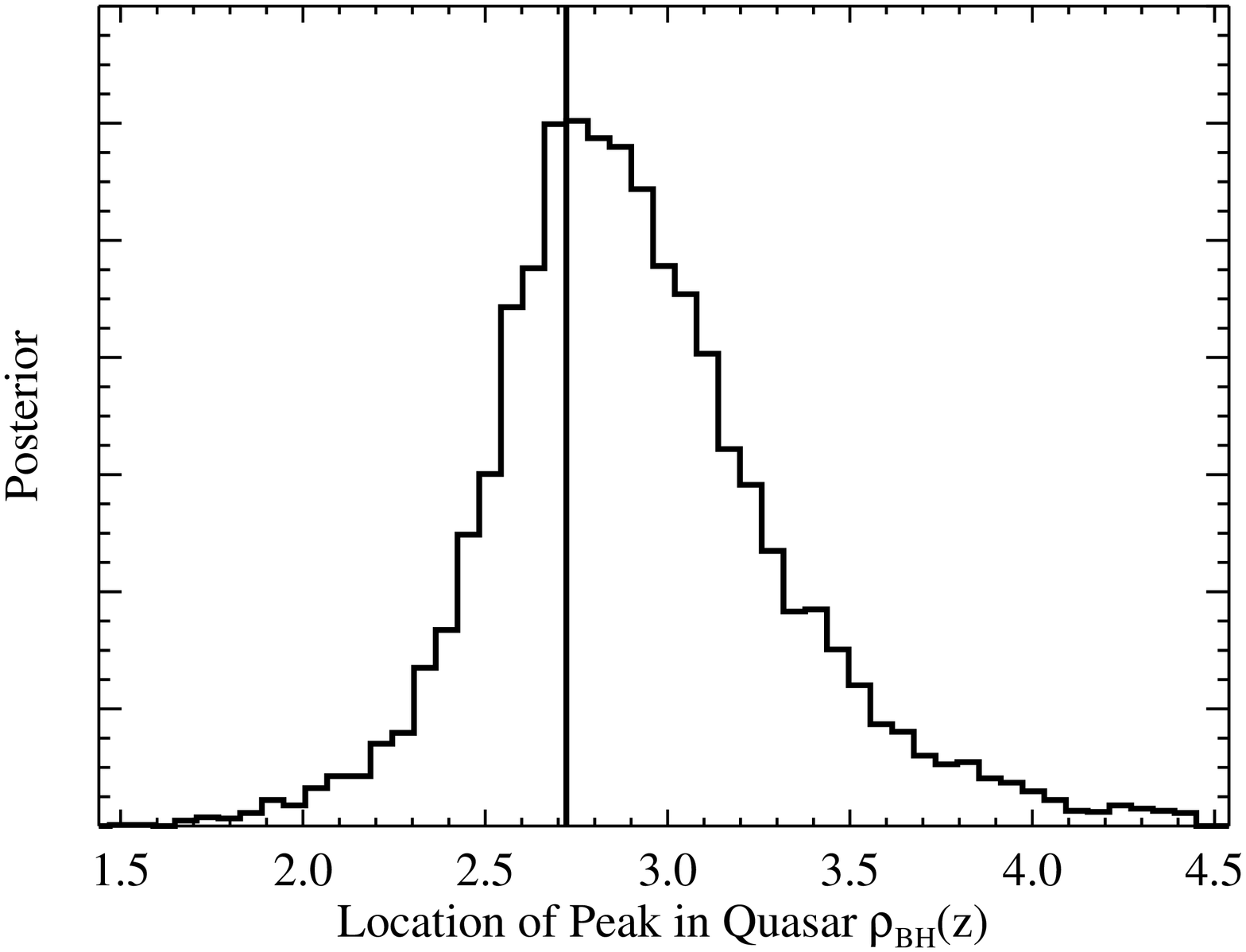}
    \caption{Posterior distribution for the redshift location of the
    peak in the comoving number density of quasars ($n(z)$, left) and
    the peak in the comoving quasar black hole mass density
    ($\rho_{BH}^{QSO}(z)$, right). The spike in the posterior at $z \approx
    0$ for values of the peak in $n(z)$ arises because the term $(dV /
    dz)^{-1}$ becomes very large at low $z$. The vertical lines denote
    the true values. The posterior distribution inferred from the MCMC
    output is able to accurately constrain the true values of the
    argumentative maximum in $n(z)$ and
    $\rho_{BH}^{QSO}(z)$.\label{f-peaks}}
  \end{center}
\end{figure}

\begin{figure}
  \begin{center}
    \includegraphics[scale=0.33,angle=90]{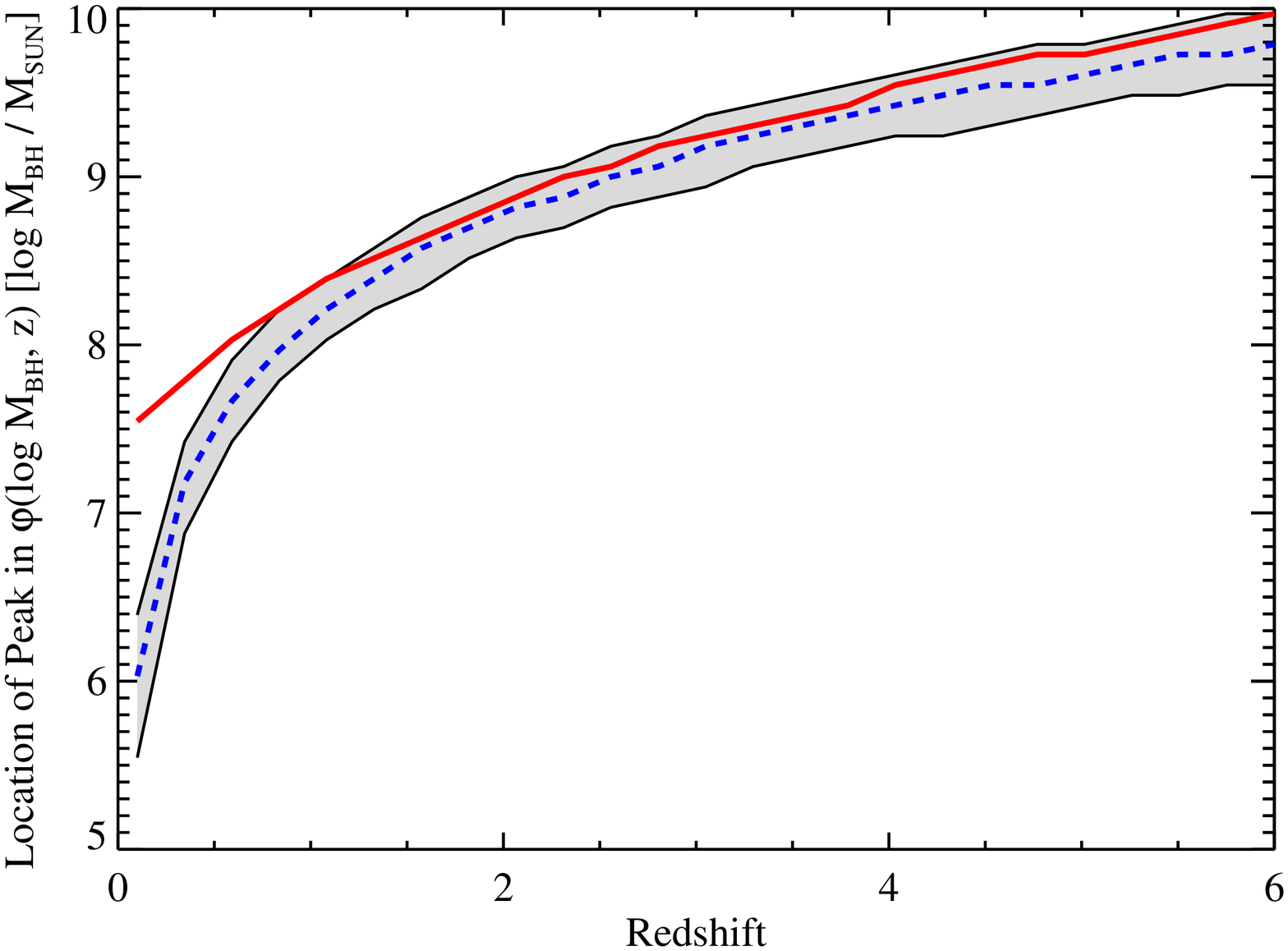}
    \includegraphics[scale=0.33,angle=90]{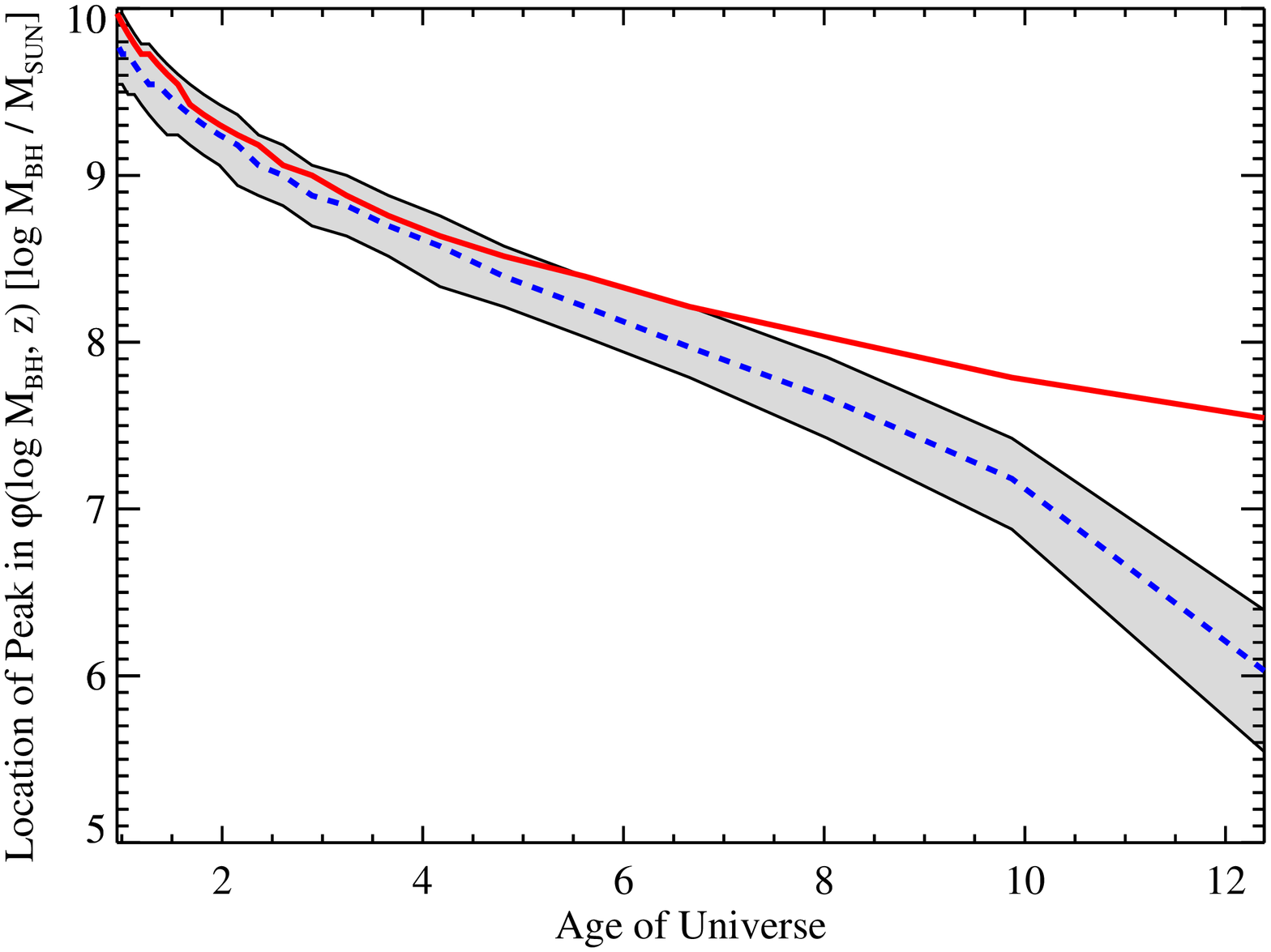}
    \caption{Location of the peak in the BHMF as a function of $z$
    (left) and cosmic age (right). The plot symbols are the same is in
    Figure \ref{f-mzmarg}. In general the posterior median of the
    Gaussian mixture model provides a good estimate of the true peak
    locations, although some bias is exhibited at the lowest
    redshifts. It is clear from these plots that the location of the
    peak in $\phi(M_{BH},z)$ evolves.\label{f-mpeakevol}}
  \end{center}
\end{figure}

  Figure \ref{f-peaks} quantifies the suggestion that $n(z)$ peaks
  later than $\rho_{BH}^{QSO}(z)$ by displaying the posterior distribution
  for the location of the respective peaks in $n(z)$ and
  $\rho_{BH}^{QSO}(z)$. While the location of the peak in $n(z)$ is highly
  uncertain we can still constrain it to be $z \lesssim 1.5$, whereas
  the location of the peak in $\rho_{BH}^{QSO}(z)$ is constrained to occur
  earlier at $2 \lesssim z \lesssim 4$. This is a consequence of the
  fact that while there were more quasars at $z \sim 1$ per comoving
  volume, their black hole masses were much higher at higher
  redshift. This evolution in characteristic $M_{BH}$ is quantified in
  Figure \ref{f-mpeakevol}, which summarizes the posterior
  distribution for the location of the peak in $\phi(\log M_{BH},z)$
  as a function of redshift and $t(z)$. As can be seen, the location
  of the peak in the BHMF shows a clear trend of increasing
  `characteristic' $M_{BH}$ with increasing $z$, although the mixture
  of Gaussian functions fit has difficulty constraining the location
  of the peak at low redshift.

  As noted in \S~\ref{s-problm}, we can use the values of $\alpha_0$
  and $\sigma_l$ to estimate the average Eddington ratio and the
  dispersion in $\log \Gamma_{Edd}$. We find $\alpha_0 =
  35.7^{+0.9}_{-1.1}, \alpha_m = 1.11^{+0.12}_{-0.10},$ and $\sigma_l
  = 0.31^{+0.06}_{-0.05}$, where the errors are at 95\%
  confidence. For a bolometric correction of $C_{\lambda} = 10$, and
  assuming that $\Gamma_{Edd}$ is independent of $M_{BH}$, this
  implies that our inferred typical Eddington ratio is $\Gamma_{Edd} =
  0.040^{+0.278}_{-0.036}$ at 95\% confidence; the estimated
  dispersion in $\log \Gamma_{Edd}$ is simply given by $\sigma_l$,
  $\sim 0.3$ dex. While the typical Eddington ratio that we infer from
  $\alpha_0$ is roughly consistent with the actual median
  $\Gamma_{Edd}$ of $0.25$, our estimated dispersion in $\Gamma_{Edd}$
  underestimates the true value of 0.5 dex. This is because we
  incorrectly assume that the $M_{BH}$--$L$ relationship is described
  by Equation (\ref{eq-problm}). Our inference regarding the Eddington
  ratio distribution is therefore biased because we assume that the
  distribution of $\Gamma_{Edd}$ does not evolve, and that the
  distribution is Gaussian. In particular, the bias resulting from the
  assumption of Gaussian dispersion appears to significantly affect
  the estimated dispersion in $\log \Gamma_{Edd}$ more than the
  estimated typical value of $\Gamma_{Edd}$, at least for our
  simulation. This is largely because the distribution in
  $\Gamma_{Edd}$ is skewed toward lower values of
  $\Gamma_{Edd}$. However, because of the flux limit, sources with low
  values of $\Gamma_{Edd}$ are undetectable. Because the dispersion in
  $\log \Gamma_{Edd}$ is estimated from the detected sources, in
  combination with the assumption of a Gaussian distribution, Equation
  (\ref{eq-problm}) is not able to pick up the additional skew at low
  $\log \Gamma_{Edd}$. As a result, the estimated dispersion in $\log
  \Gamma_{Edd}$ is underestimated when assuming a Gaussian
  distribution. We note that this bias is not a feature of our
  algorithm, but affects any analysis that attempts to infer the
  distribution of Eddington ratios using a flux-limited sample.

\begin{figure}
  \begin{center}
    \includegraphics[scale=0.33,angle=90]{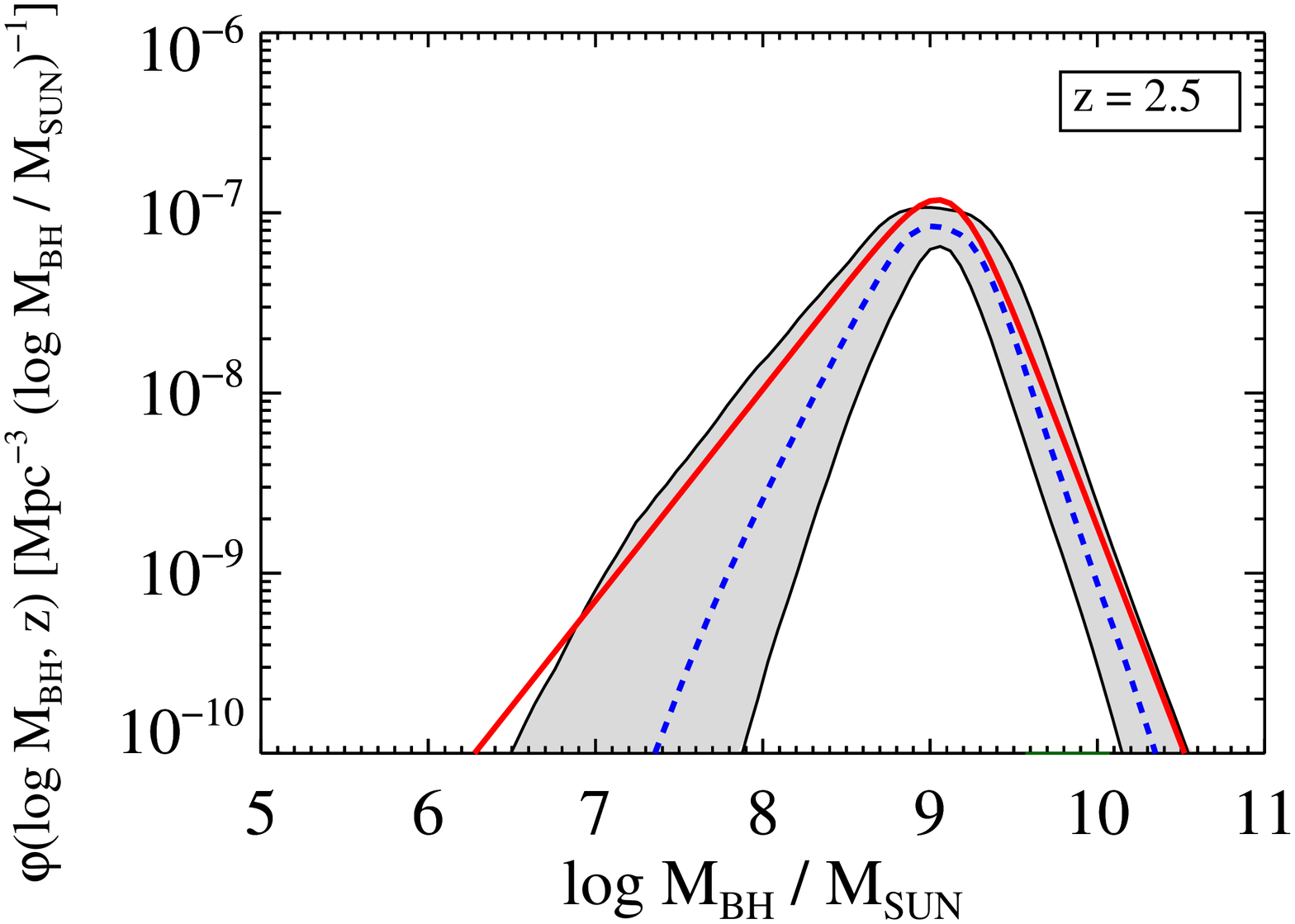}
    \includegraphics[scale=0.33,angle=90]{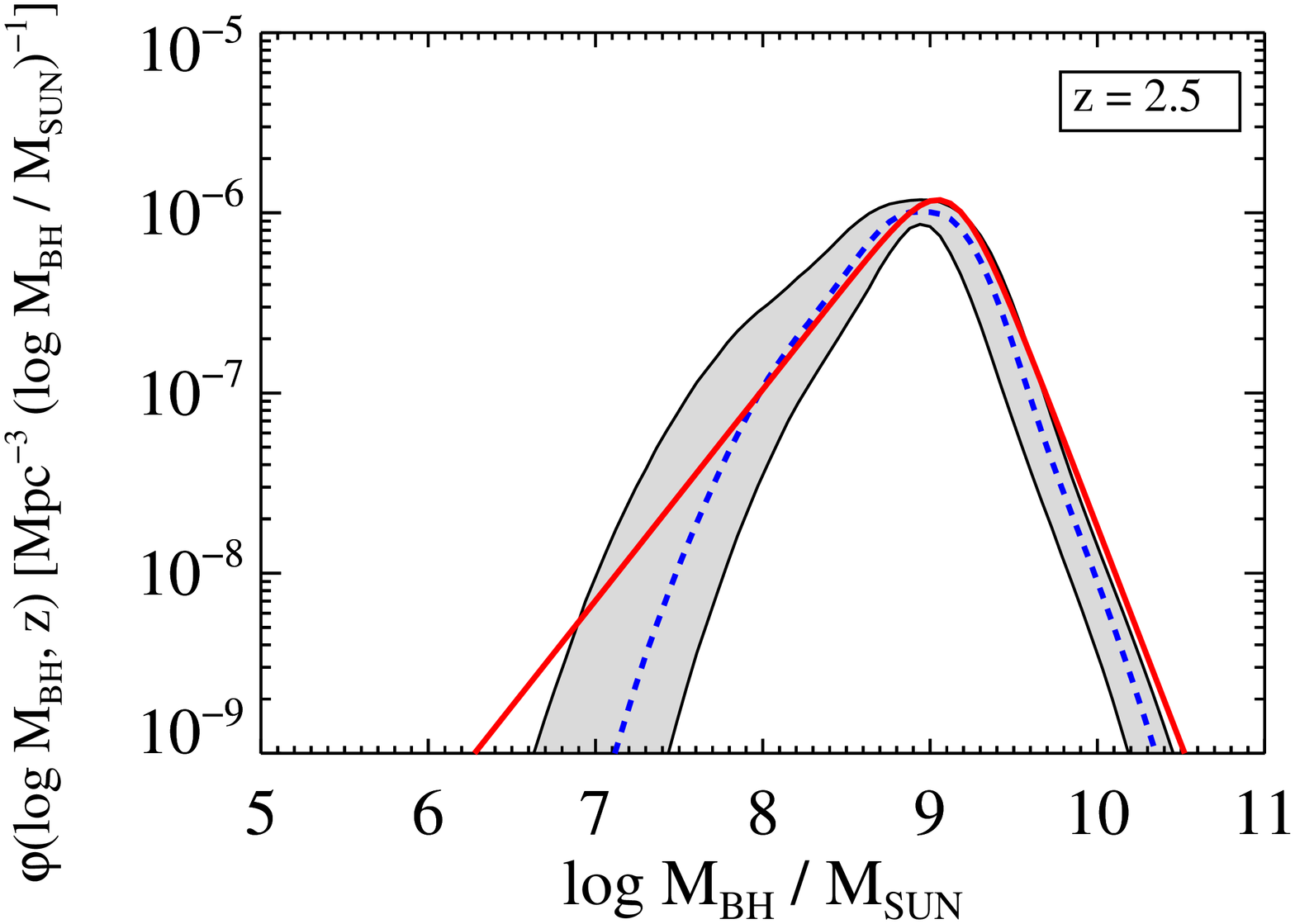}
    \caption{BHMF at $z = 2.5$ for the simulated sample with $n \sim
    1000$ detected sources (left) and $n \sim 10^4$ detected sources
    (right); the left panel is the same as the $z = 2.5$ BHMF shown in
    Figure \ref{f-philog}. The uncertainties derived for the $n \sim
    10^4$ are smaller than for the $n \sim 1000$ sample, particularly
    at low $M_{BH}$ where the survey becomes incomplete. However, the
    uncertainties for the $n \sim 10^4$ survey at high $M_{BH}$, where
    the survey is complete, are not considerably smaller than those
    for the $n \sim 1000$ survey. This is because the BHMF estimate is
    limited by the systematic uncertainty in the broad line mass
    estimate normalization, derived from $\beta_0$, and the broad line
    mass estimate statistical error, derived from
    $\sigma_{BL}$. Because the observed distribution of luminosities
    and line widths does not convey any information on these two
    quantities, increasing the sample size will not reduce the
    uncertainty on the BHMF beyond the systematic uncertainty on
    $\beta_0$ and $\sigma_{BL}$. \label{f-bhmfcompare}}
  \end{center}
\end{figure}

  In order to assess how the inferred BHMF depends on the sample size,
  we simulated a second data set in the sammer manner as described
  above, but used $N = 2 \times 10^6$ sources for the BHMF
  normalization. This gave us $n \sim 10^4$ detected quasars. In
  Figure \ref{f-bhmfcompare} we compare the estimated BHMF at $z =
  2.5$ for the survey with $n \sim 1000$ sources and $n \sim 10^4$
  sources. The uncertainties are lower for the survey with more
  sources, where the most noticeable improvement occurs at low
  $M_{BH}$. However, the increased sample size did not offer a
  significant amount of improvement at high $M_{BH}$, where sources
  are more easily detected. This is likely because the uncertainty in
  the broad line mass estimate normalization, $\beta_0$, and intrinsic
  scatter, $\sigma_{BL}$, dominates the uncertainty in the BHMF at
  high $M_{BH}$. Because we cannot constrain $\beta_0$ and
  $\sigma_{BL}$ from the distribution of line widths and luminosities,
  the data do not contain any information on $\beta_0$ and
  $\sigma_{BL}$. Therefore, the likelihood function is unable to
  convey any information on $\beta_0$ and $\sigma_{BL}$, and all of
  the information comes from the prior distribution. As a result, our
  ability to constrain the BHMF is limited by the statistical
  uncertainty on $\beta_0$ and $\sigma_{BL}$, and an increase in the
  sample size will eventually not result in a decrease in the
  uncertainty on the BHMF. The only way to reduce the uncertainty on
  the BHMF for large surveys is to better constrain the broad line
  mass estimate normalization and statistical uncertainty, most likely
  by increasing the sample of AGN with reverberation mapping data.

  Throughout this work we have assumed that the selection function is
  known, and that $\beta_0$, and $\sigma_{BL}$ are known within some
  statistical uncertainty.  However, this may not be the case, and
  before concluding this section we briefly discuss how systematic
  error in the selection function, $\beta_0$ and $\sigma_{BL}$, affect
  the inferred BHMF. We did not experiment with incorrect selection
  functions, and so it is not entirely clear how robust BHMF
  estimation is to errors in the selection function. However, from
  Equations (\ref{eq-thetapost}) and (\ref{eq-npost}) it is clear that
  the selection function only enters into the posterior probability
  distribution (or likelihood function) via an integral that averages
  the selection function over the joint distribution of luminosity and
  redshift (i.e., the luminosity function). As a result, errors in the
  selection function will be smoothed out. Furthermore, they will be
  suppressed in regions where values of the luminosity function are
  small. Based on this, we do not think it likely that small errors in
  the selection function will introduce significant bias into the
  results; however, if the errors in the selection function are large
  enough to significantly bias the value of $p(I=1|L,z)$, then the
  results may be significantly biased as well.

  It is useful to work directly with the broad line mass estimates to
  assess the effect that systematic uncertainty on the values of the
  broad line mass estimate normalization and statistical uncertainty
  have on the inferred BHMF. Ignoring selection effects, one can think
  of our method as `correcting' the BHMF inferred from binning up the
  broad line mass estimates. Therefore, if $\beta_0$ is systematically
  underestimated, then this will result in a shift of the inferred
  BHMF toward higher masses. Similarly, if $\beta_0$ is systematically
  overestimated, than the inferred BHMF will be shifted toward lower
  masses. In addition, the value of $\sigma_{BL}$ controls how much
  the BHMF inferred from the broad line mass estimates is artificially
  broadened by the statistical uncertainty in $\hat{M}_{BL}$. A higher
  value of $\sigma_{BL}$ will result in a greater amount of
  broadening. Therefore, if our assumed values of $\sigma_{BL}$ are
  systematically overestimated, then we would infer a greater amount
  of broadening than is real. As a result, our correction would be too
  large, and we would infer an intrinsic BHMF that is too
  narrow. Similarly, if our assumed values of $\sigma_{BL}$ are
  systematically underestimated, then we would not correct enough for
  the statistical uncertainty in the broad line mass estiamtes, and we
  would infer an intrinsic BHMF that is too broad.

  \subsection{Using the MCMC Output to Evaluate the BHMF Fit}

  \label{s-postcheck}

  Throughout this section we have been analyzing the MCMC results by
  comparing to the true BHMF. However, in practice we do not have
  access to the true BHMF, and thus a method is needed for assessing
  the quality of the fit. As in KFV08, the statistical model may be
  checked using a technique known as posterior predictive checking
  \citep[e.g.,][]{rubin81,rubin84,gelman98}. Here, the basic idea is
  to use each of the MCMC outputs to simulate a new random observed
  data set. The distributions of the simulated observed data sets are
  then compared to the true observed data in order to assess whether
  the statistical model gives an accurate representation of the
  observed data. It is important to construct simulated data sets for
  each of the MCMC draws in order to incorporate our uncertainty in
  the model parameters.

  Random draws for $M_{BH}$ and $z$ for each MCMC draw may be obtained
  according to the procedure outlined in \S~7.3 of KFV08, after
  replacing $L$ with $M_{BH}$. Once one obtains a random draw of
  $M_{BH}$ and $z$, simulated values of $L_{\lambda}$ may be obtained
  using Equation (\ref{eq-problm}) with $\alpha_0, \alpha_m,$ and
  $\sigma_l$. Then, given these values of $L_{\lambda}$ and $M_{BH}$,
  values of $v$ for each emission line can be simulated from Equation
  (\ref{eq-probvlm}) using the values of $\beta_0, \beta_l,$ and
  $\sigma_{BL}$. Simulation from Equation (\ref{eq-probvlm}) requires
  a value of $\alpha_{\lambda}$ in order to convert $L_{\lambda}$ to
  $L_{\lambda}^{BL}$. In order to account for the range in continuum
  slopes, we randomly draw of value of $\alpha_{\lambda}$ from our
  data set and use this value to convert to $L_{\lambda}^{BL}$. These
  simulated values of $L_{\lambda}, z,$ and ${\bf v}$ are then folded
  through the selection function, leaving one with a simulated
  observed data set $({\bf v}_{obs}, L_{obs}, z_{obs})$. This process
  is repeated for all values of $N$ and $\theta$ obtained from the
  MCMC output, leaving one with simulated observed data sets of $({\bf
  v}_{obs}, L_{obs}, z_{obs})$. These simulated observed data sets can
  then be compared with the true distribution of ${\bf v}_{obs},
  L_{obs},$ and $z_{obs}$ to test the statistical model for any
  inconsistencies.

\begin{figure}
  \begin{center}
    \scalebox{0.5}{\rotatebox{90}{\plotone{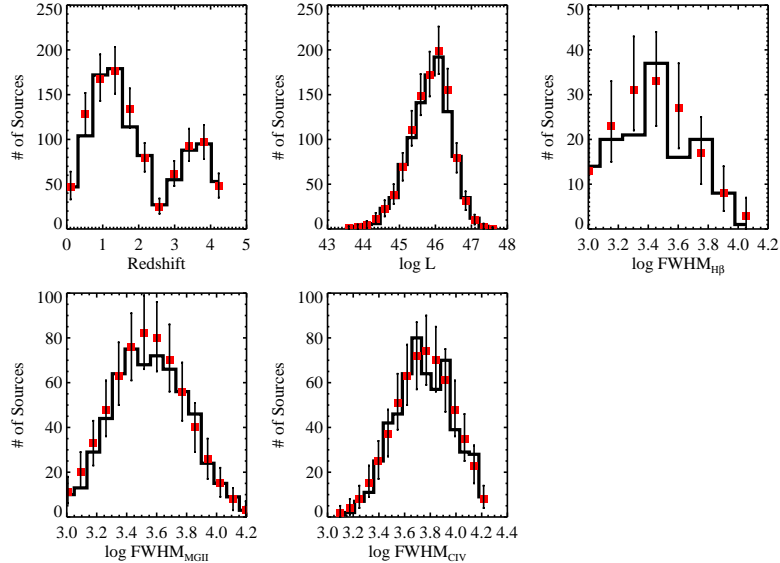}}}
    \caption{Posterior predictive check for the Gaussian mixture model
    (see \S~\ref{s-postcheck}). The histograms show the actual
    distributions of $\log L_{obs},z_{obs},$ and $\log FWHM_{obs}$,
    the red squares denote the posterior medians for the number of
    sources in each respective bin, and the error bars contain the
    inner $90\%$ of the histogram values for the samples simulated
    from the posterior. The mixture of Gaussian functions model is
    able to provide an accurate prediction of the observed
    distribution of luminosity, redshift, and line widths, and thus
    there is not any evidence to reject it as providing a poor
    fit. \label{f-postcheck}}
  \end{center} 
\end{figure}

  In Figure \ref{f-postcheck} we show histograms for the observed
  distributions of $z$, $\log L_{\lambda}$, and $\log FWHM$ for the
  H$\beta$, Mg II, and C IV emission lines. These histograms are
  compared with the posterior median of the observed distributions
  based on the mixture of Gaussian functions model, as well as error
  bars containing $90\%$ of the simulated observed values. As can be
  seen, the distributions of the observed data sets simulated from our
  assumed statistical model are consistent with the distributions of
  the true observed data, and therefore there is no reason to reject
  the statistical model as providing a poor fit.

  \section{APPLICATION TO BQS QUASARS}

  \label{s-bqs}

  As a final illustration of our method we used it to estimate the low
  redshift active BHMF from the 87 $z < 0.5$ quasars from the Bright
  Quasar Survey \citep[BQS,][]{bqs}. The H$\beta$ line widths and
  continuum luminosities for 71 of the BQS quasars are taken from
  Table 7 of \citet{vest06}, and 16 of the quasars in the
  \citet{boroson92} sample have black hole mass estimates from
  reverberation mapping \citep{peter04}. For each source with
  reverberation mapping data, we used the first entry of $\lambda
  L_{\lambda}(5100$\AA$)$ in Table 1 of \citet{vest06} as the
  single-epoch luminosity; these values were based on continuum
  luminosities reported by \citet{boroson92} or \citet{marz03}. We
  assumed measurement errors of $10\%$ on the emission line
  $FWHM$. The BQS sample covers an area of $\Omega = 10,714\ {\rm
  deg}^2$ and is selected with an average flux limit of $B = 16.16$
  \citep{bqs}, with no apparent correlation with redshift and $U - B$
  color \citep{jester05}. We converted the $B = 16.16$ flux limit to a
  flux limit at $5100$\AA\ assuming a power law continuum, $f_{\nu}
  \propto \nu^{-\alpha}$, with $\alpha = 0.5$ \citep{richards01}. We
  used $K = 3$ Gaussian functions to fit $\phi(M_{BH},z)$ for $z <
  0.5$.

  Because we are including the actual values of $M_{BH}$ for the 16
  reverberation mapping sources, the contribution to the posterior for
  these sources is
  \begin{equation}
    p(\theta|M_{BH},L_{\lambda},z) = \prod_{i=1}^{16} p(\log L_{\lambda,i}|M_{BH,i},\theta) 
    p(\log M_{BH,i},\log z_i|\theta).
    \label{eq-reverbpost}
  \end{equation}
  Here, $p(L_{\lambda,i}|M_{BH,i},\theta)$ is given by Equation
  (\ref{eq-problm}) and $p(\log M_{BH,i},\log z_i|\theta)$ is given by
  Equation (\ref{eq-mixmod}). The product in Equation
  (\ref{eq-reverbpost}) is only over the quasars with $M_{BH}$
  estimated from reverberation mapping, whereas the contribution to
  the posterior for the BQS sources without reverberation mapping is
  given by Equation (\ref{eq-thetapost}). The posterior for the
  complete BQS sample is then the product of Equation
  (\ref{eq-reverbpost}) and Equation (\ref{eq-thetapost}).

\begin{figure}
  \begin{center}
    \includegraphics[scale=0.33,angle=90]{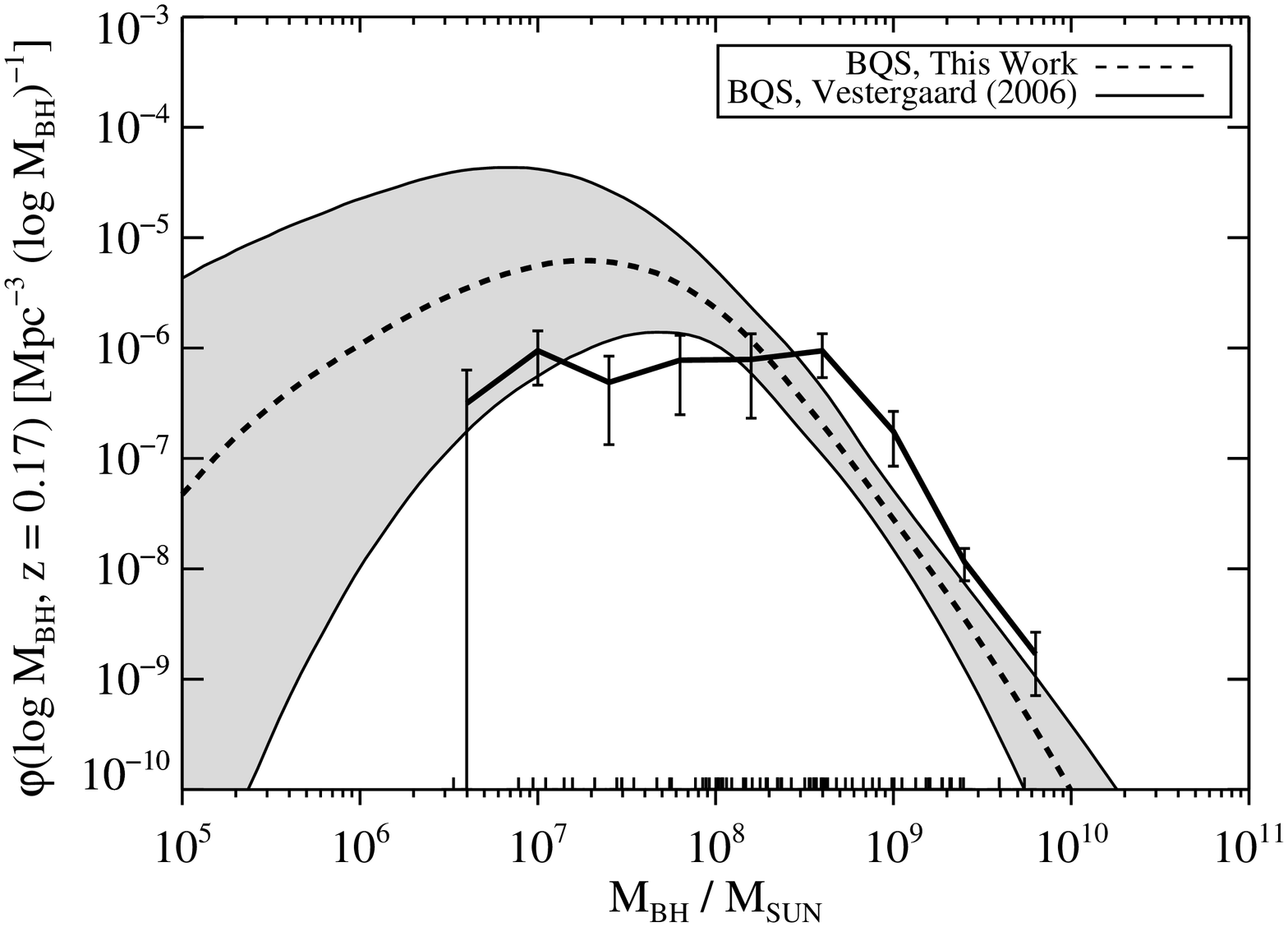}
    \includegraphics[scale=0.33,angle=90]{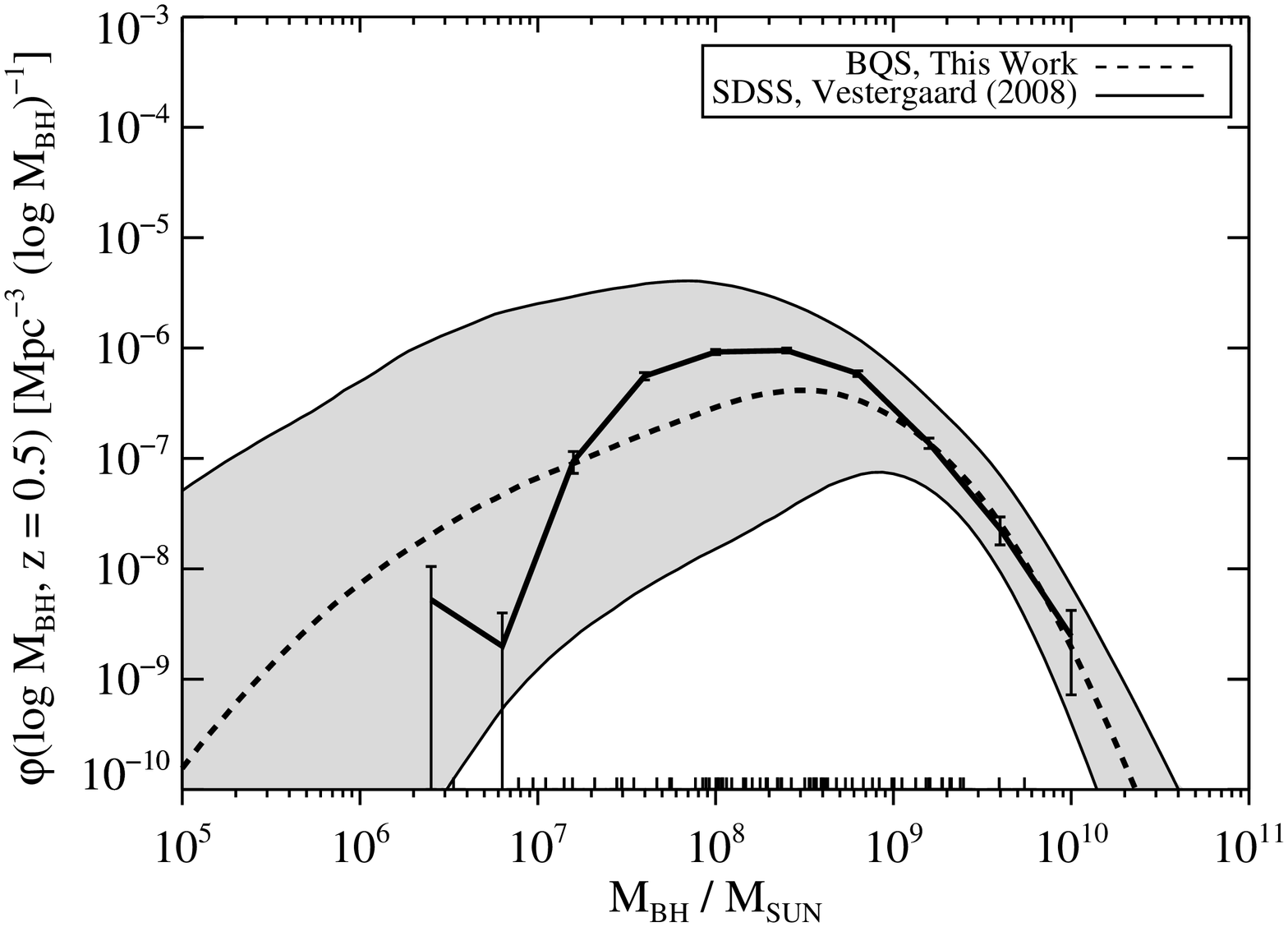}
    \caption{The $z = 0.17$ (left) and $z = 0.5$ (right) broad line
    quasar BHMF as estimated from the BQS sample. The dashed line
    denotes the posterior median for the mixture of Gaussian functions
    model, the shaded region contains $68\%$ of the posterior
    probability, and the tick marks along the $x$-axis mark the
    locations of the broad line mass estimates. The estimate of the $z
    = 0.17$ BHMF becomes significantly uncertain at $M_{BH} \lesssim
    10^8 M_{\odot}$, and the $z = 0.17$ BHMF appears to fall off as a
    power law above $M_{BH} \gtrsim 10^8 M_{\odot}$. The $z = 0.5$
    BHMF is not very well constrained, but there is evidence for a
    shift in the BHMF toward higher $M_{BH}$ from $z = 0.17$ to $z =
    0.5$. For comparison, we show the BHMF estimated by
    \citet{vestbqs} using the BQS sources (left, solid line with error
    bars), and the BHMF estimated by \citet{vest08} using the SDSS DR3
    quasars (right, solid line with error bars). The shift in the BHMF
    inferred from the binned mass estimates is apparent in the BQS
    sample, while the SDSS and BQS $z = 0.5$ BHMF estimates agree
    fairly well.
    \label{f-bqsbhmf}}
  \end{center}
\end{figure}

  In Figure \ref{f-bqsbhmf} we show the $z = 0.17$ BHMF derived from
  the BQS sample. Also shown is the binned BHMF for the BQS sources,
  calculated directly from the broad line mass estimates by
  \citet{vestbqs}. We show the BHMF at $z = 0.17$ because the average
  redshift of the BQS sources is $z \approx 0.17$, therefore allowing
  a more direct comparison between the binned BHMF and the BHMF
  derived using our mixture of Gaussian functions approach. In
  addition, the uncertainties on our estimated BHMF are smallest at $z
  \approx 0.17$. We are able to place some constraints on the local
  BHMF, despite the fact that the BQS sample only contains 87 sources
  and has a very shallow flux limit. The $z \sim 0.2$ quasar BHMF
  appears to fall off as a power law above $M_{BH} \gtrsim 10^8
  M_{\odot}$. Unfortunately, our estimate of the local BHMF becomes
  considerably uncertain below $M_{BH} \lesssim 10^8 M_{\odot}$, so it
  is unclear to what degree the power law trend continues below this
  point. In addition, the binned estimate overestimates the BHMF at the
  high $M_{BH}$ end due to the intrinsic uncertainty in the broad line
  mass estimates, and underestimates the BHMF at the low $M_{BH}$ end
  due to incompleteness, in agreement with our simulations (see
  \S~\ref{s-simmcmc}).

  In Figure \ref{f-bqsbhmf} we also compare our estimate of the BHMF
  at $z = 0.5$ with the $z = 0.5$ BHMF as reported by
  \citet{vest08}. \citet{vest08} estimated the $z = 0.5$ BHMF by
  binning estimates of $M_{BH}$ derived from the H$\beta$ and Mg II
  broad emission lines over the redshift range $0.3 < z < 0.68$, using
  the SDSS DR3 quasar catalogue \citep{dr3qsos}. Despite the
  differences in approach and survey selection, the two estimates of
  the $z = 0.5$ BHMF agree fairly well. However, because $z = 0.5$
  defines the upper redshift limit of our BQS sample, the
  uncertainties on the BHMF derived from the BQS quasars are very
  large. In addition, incompleteness in $M_{BH}$ likely affects the
  low $M_{BH}$ bins of the \citet{vest08}, causing the \citet{vest08}
  $z = 0.5$ BHMF to underestimate the true $z = 0.5$ BHMF in these
  bins, a fact reflected by the larger error bars. However, a direct
  comparison between our Bayesian approach and the \citet{vest08}
  estimate is difficult, due to the different redshift ranges used to
  estimate the BHMF, and the different selection methods of the BQS
  and the SDSS.

\begin{figure}
  \begin{center}
    \scalebox{0.5}{\rotatebox{90}{\plotone{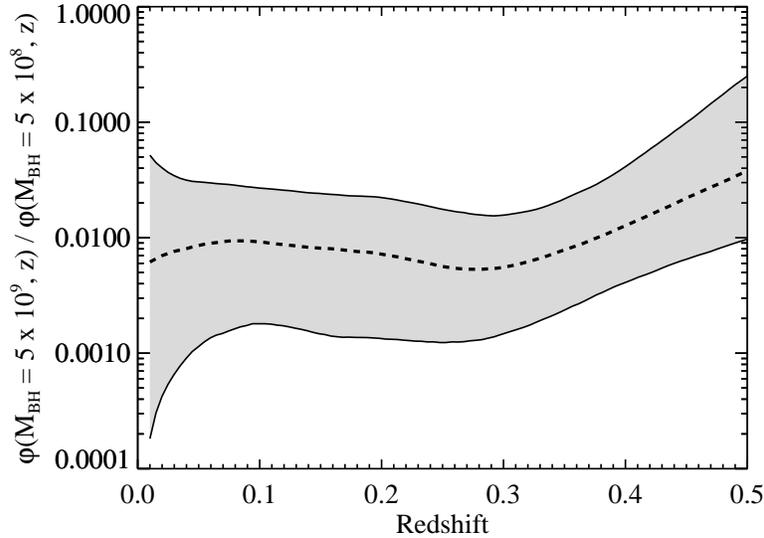}}}
    \caption{The ratio of the broad line quasar BHMF at $M_{BH} = 5
    \times 10^9 M_{\odot}$ compared to the BHMF at $M_{BH} = 5 \times
    10^8 M_{\odot}$, as a function of $z$ and estimated from the BQS
    quasars. The dashed line is the posterior median, and the shaded
    region contains 68\% of the probability. Assuming that the BHMF is
    a power-law from $M_{BH} = 5 \times 10^8 M_{\odot}$ to $M_{BH} = 5
    \times 10^9 M_{\odot}$, the logarithm of this ratio is the slope
    of the BHMF. The high $M_{BH}$ BHMF slope appears to be fairly
    constant for $z \lesssim 0.3$ with a slope of $\sim 2$, and there
    is marginal evidence for a flattening of the high $M_{BH}$ slope
    at $z \gtrsim 0.3$.
    \label{f-bqsmax}}
  \end{center}
\end{figure}

  Although the BQS has a small sample size and probes a narrow range
  in $z$, we can attempt to quantify any evolution in the local BHMF
  by comparing the ratio of the comoving number density of quasars at
  two different values of $M_{BH}$. Comparison of the estimated BHMF
  at $z = 0.17$ and $z = 0.5$ suggests a shift in the BHMF toward
  large $M_{BH}$. In Figure \ref{f-bqsmax} we show the best fit values
  of the ratio of $\phi(\log M_{BH},z)$ at $M_{BH} = 5 \times 10^8
  M_{\odot}$ to $\phi(\log M_{BH},z)$ at $M_{BH} = 5 \times 10^9
  M_{\odot}$ as a function of $z$, as well as the $68\%$ confidence
  interval. The logarithm of this ratio gives the slope of a power-law
  between $M_{BH} = 5 \times 10^{8} M_{\odot}$ and $M_{BH} = 5 \times
  10^9 M_{\odot}$, and therefore allows us to probe evolution in the
  shape of the quasar BHMF at the high $M_{BH}$ end. In general, the
  ratio is fairly flat, implying no evolution in the high $M_{BH}$
  slope of the BHMF. However, at $z \gtrsim 0.3$ there is marginal
  evidence for a flattening of the high $M_{BH}$ slope of the
  BHMF. The values of this ratio imply that the BHMF at the high
  $M_{BH}$ end falls off as a power-law with slope $\sim 2$, although
  slopes of $\sim 1$ and $\sim 3$ are also consistent with the BQS
  quasars.

\begin{figure}
  \begin{center}
    \scalebox{0.5}{\rotatebox{90}{\plotone{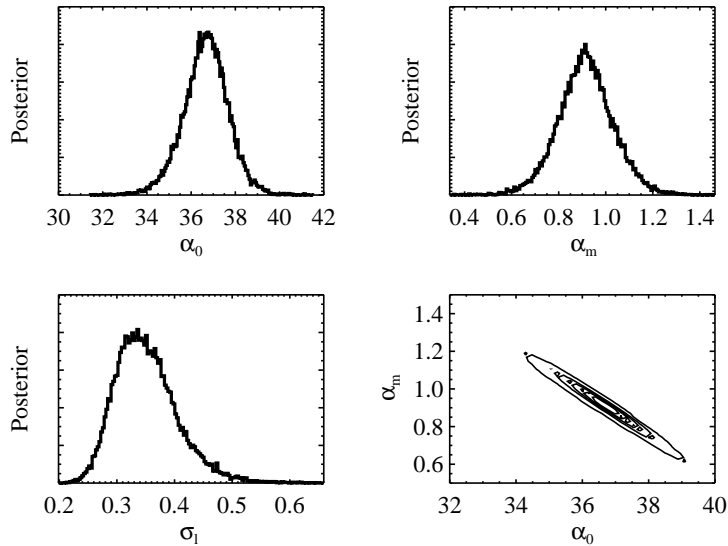}}}
    \caption{Posterior distributions of the parameters for the
    distribution of luminosities at a given $M_{BH}$, as estimated
    from the $z < 0.5$ BQS quasars. The uncertainty on $\alpha_0$ and
    $\alpha_m$ is highly correlated. Assuming a bolometric correction
    of $C_{5100} \sim 10$, the values of $\alpha_0$ and $\sigma_l$
    imply that the $z < 0.5$ distribution of broad line quasar
    Eddington ratios has a mean of $\Gamma_{Edd} \sim 0.4$ and a
    dispersion of $\sim 0.5$ dex. \label{f-bqspost}}
  \end{center}
\end{figure}

  In figure \ref{f-bqspost} we summarize the posterior probability
  distribution for the parameters governing the distribution of
  $L_{\lambda}$ at a given $M_{BH}$ (see Eq.[\ref{eq-problm}]). Based
  on the MCMC results, we can constrain the $M_{BH}$--$\lambda
  L_{\lambda}(5100\AA)$ relationship at $z < 0.5$ to be
  \begin{equation}
    \lambda L_{\lambda} (5100\AA) = 5.18^{+429}_{-5.14} \times 10^{36} 
    \left(\frac{M_{BH}}{M_{\odot}} \right)^{0.92 \pm 0.24}\ [{\rm erg\ s^{-1}}],
    \label{eq-bqsml}
  \end{equation}
  where we have quoted the errors at $95\%$ confidence. The dispersion
  in $L_{5100}$ at a given $M_{BH}$ is estimated to be $\sigma_l =
  0.35^{+0.13}_{-0.08}$. Assuming that the bolometric correction is on
  average $C_{5100} \sim 10$ \citep[e.g.,][]{kaspi00}, comparison of
  Equation (\ref{eq-bqsml}) with Equation (\ref{eq-mlrel}) suggests
  that $z < 0.5$ broad line AGN have typical Eddington ratios of
  $\Gamma_{Edd} \sim 0.4$. As argued in \S~\ref{s-problm}, the
  distribution in $\log L_{\lambda}$ at a given $M_{BH}$ is the
  convolution of the distribution of $\log \Gamma_{Edd}$ with the
  distribution of $\log C_{\lambda}$. Therefore, the dispersion in
  $L_{\lambda}$ at a given $M_{BH}$ is a combination of the dispersion
  in Eddington ratio and bolometric correction. As a result, we are
  unable to estimate the dispersion in Eddington ratios at a given
  $M_{BH}$ from $\sigma_l$. However, if the bolometric correction to
  $L_{5100}$ increases with increasing Eddington ratio, as found by
  \citet{vasud07}, or if the bolometric correction is independent of
  $\Gamma_{Edd}$, then the dispersion in $\Gamma_{Edd}$ must be less
  than $\sigma_l$. Therefore, because we infer that $\sigma_l \lesssim
  0.5$ dex, our results imply that the dispersion in Eddington ratios
  at a given $M_{BH}$ is $\lesssim 0.5$ dex for $z < 0.5$ broad line
  quasars. These results on the Eddington ratio distribution are
  consistent with previous work
  \citep[e.g.,][]{mclure04,vest04,koll06}; however, they may be biased
  because of our assumption of a Gaussian and non-evolving Eddington
  ratio distribution. In particular, if the distribution of Eddington
  ratios is skewed toward low $\log \Gamma_{Edd}$, then we will have
  underestimated the intrinsic dispersion in $\log \Gamma_{Edd}$.

  \section{SUMMARY}
  
  \label{s-summary}

  We have derived the observed data likelihood function which relates
  the quasar BHMF to the observed distribution of redshifts,
  luminosities, and broad emission line widths. This likelihood
  function is then used in a Bayesian approach to estimating the BHMF,
  where the BHMF is approximated as a mixture of Gaussian
  functions. Because much of this work was mathematically technical,
  we summarize the important points here.
  \begin{itemize}
  \item
    In this work we describe a flexible parameteric model for the
    BHMF, where the BHMF is modeled as a mixture of Gaussian
    functions. The distribution of luminosities is modelled as a
    linear regression of $\log L_{\lambda}$ as a function of $\log
    M_{BH}$, where the distribution of $\log L_{\lambda}$ at a given
    $M_{BH}$ was assumed to follow a normal distribution. The
    distribution in line widths at a given $L_{\lambda}$ and $M_{BH}$
    is also assumed to have the form of a linear regression, where the
    parameters are based on the most recent broad line mass
    estimates. Equation (\ref{eq-mixbhmf}) gives the BHMF under the
    mixture of Gaussian function model.

    Equations (\ref{eq-mixlik1}) and (\ref{eq-mixdetprob}) define the
    likelihood function for broad line mass estimates under the
    mixture of Gaussian functions model if only one emission line at a
    given $z$ is used to estimate $M_{BH}$. Otherwise, if multiple
    emission lines are used for a single quasar, then Equation
    (\ref{eq-mixlik2}) must be used. The posterior is then found by
    inserting the prior distribution and likelihood function into
    Equations (\ref{eq-thetapost}) and (\ref{eq-npost}).
  \item
    Using methods developed for luminosity function estimation (e.g.,
    $1 / V_a$-type estimators) without modification will lead to
    errors in black hole mass function estimation, as the black hole
    mass selection function is not equivalent to the flux selection
    function. In addition, using broad line estimates of $M_{BH}$ will
    lead to a broader inferred BHMF if one does not correct for the
    intrinsic uncertainty in the broad line mass estimates. This
    causes one to overestimate $\phi(M_{BH},z)$ in the tails of the
    distribution, and underestimate $\phi(M_{BH},z)$ near the peak of
    the distribution. However, because low $M_{BH}$ AGN are more
    likely to be missed by flux-limited surveys, $\phi(M_{BH},z)$ will
    be underestimated at low $M_{BH}$ due to incompleteness. The end
    result is a spurious shift in the inferred BHMF toward higher
    $M_{BH}$: incompleteness at low $M_{BH}$ causes one to miss low
    $M_{BH}$ sources while the intrinsic statistical uncertainty on
    the broad line mass estimates causes one to overestimate the
    number of high $M_{BH}$ black holes.
  \item
    In \S~\ref{s-measerr} we modify the likelihood function to include
    measurement error in the emission line width. We show that if the
    measurement errors on the line width are much smaller than the
    intrinsic physical dispersion in line widths, then measurement
    error may be neglected. However, if measurement error on the line
    width is a concern, Equations
    (\ref{eq-probvlz2})--(\ref{eq-cprobv}) should be used for Equation
    (\ref{eq-probvlz}) instead of Equation (\ref{eq-mixlik1}).
  \item
    We describe in \S~\ref{s-mha} a Metropolis-Hastings algorithm
    (MHA) for obtaining random draws from the posterior distribution
    of the BHMF under the mixture of Gaussian functions model. These
    random draws may be used to estimate the posterior distribution
    for the BHMF, as well as to estimate the posterior for any
    quantities calculated from the BHMF. The posterior provides
    statistically accurate uncertainties on the BHMF and related
    quantities, even below the survey detection limits. We use
    simulation in \S~\ref{s-sim} to illustrate the effectiveness of
    our statistical method, as well as to give an example on how to
    use the MHA output to perform statistical inference.
  \item
    We concluded by applying our method to obtain an estimate of the
    local unobscured quasar BHMF from the $z < 0.5$ BQS quasar
    sample. Although there is little information in the BQS quasars on
    the BHMF at $M_{BH} \lesssim 10^8 M_{\odot}$, the mixture of
    Gaussian functions estimate suggests that the local quasar BHMF
    falls off approximately as a power law with slope $\sim 2$ for
    $M_{BH} \gtrsim 10^8 M_{\odot}$ at $z \approx 0.2$. The
    local quasar BHMF appears to shift toward larger $M_{BH}$ at
    higher $z$, and there is marginal evidence for a flattening of the
    high mass BHMF slope at $z \gtrsim 0.3$. We estimate that at a
    given $M_{BH}$, $z < 0.5$ broad line quasars have a typical
    Eddington ratio of $\sim 0.4$ and a dispersion in Eddington ratio
    of $\lesssim 0.5$ dex. However, the estimate of the dispersion in
    Eddington ratio could be biased toward smaller values if the true
    distribution of Eddington ratios is significantly skewed toward
    lower values.
  \end{itemize}

  \acknowledgements

  BCK, XF and MV acknowledge support for NSF grants AST 03-07384,
  08-06861 and a Packard Fellowship for Science and Engineering. BK
  acknowledges support by NASA through Hubble Fellowship grant
  \#HF-01220.01 awarded by the Space Telescope Science Institute,
  which is operated by the Association of Universities for Research in
  Astronomy, Inc., for NASA, under contract NAS 5-26555. We also
  acknowledge financial support from HST grants HST-GO-10417 (XF, MV),
  HST-AR-10691 (MV), and HST-GO-10833 (MV) awarded by the Space
  Telescope Science Institute, which is operated by the Association of
  Universities for Research in Astronomy, Inc., for NASA, under
  contract NAS5-26555.

\end{document}